\documentclass[12pt]{article}

\usepackage{amsmath}
\usepackage{amssymb}
\usepackage{latexsym}
\usepackage{theorem}
\usepackage{xspace}
\usepackage{epsfig}
\usepackage{infernce}
\usepackage{epic,ecltree}

\usepackage{url}

\renewcommand{\predicate}[1]{$ #1 $ }

\newcommand{\Tree}{\mathsf{T}} 
\newcommand{\vtr}{\vartriangleright}

\def\dlb{\mathopen{[\kern-1.4pt[}}
\def\drb{\mathclose{]\kern-1.4pt]}}

\def\wedgevee{\ensuremath{\mathbin{\rlap{$\vee$}\wedge}}}
\def\boxdiamond{\ensuremath{\mathbin{\rlap{$\Box$}\kern.5pt\raisebox{.5pt}{$\lozenge$}}}}
\def\select{\mathsf{select}}

\def\K{\ensuremath{\mathsf{K}}}
\def\vtrstar{\mathbin{\vtr\!\!\!^*}}
\def\vdstar{\mathbin{\vdash\!\!^*}}
\def\vdsstar{\mathbin{\vdash\!\!\!_\succ\!\!\!^*}}
\def\vds{\mathbin{{\vdash\!\!\!_\succ}}}
\def\vdastar{\mathbin{\vdash\!\!\!_{\textit{ax}}\!\!\!^*}}
\def\vda{\mathbin{{\vdash\!\!\!_{\textit{ax}}}}}

\def\mfA{\mathfrak{A}}
\def\bfq{\mathbf{q}}

\def\mcP{\mathcal{P}}
\def\mcM{\mathcal{M}}
\def\mcS{\mathcal{S}}
\def\mcT{\mathcal{T}}

\def\N{\mathbb{N}}

\def\EXPTIME{\textsc{ExpTime}}
\def\NEXPTIME{\textsc{NExpTime}}

\newcommand{\qed}{{\unskip\nobreak\hfil\penalty50
   \hskip2em\hbox{}\nobreak\hfil
   \qedd
   \parfillskip=0pt \finalhyphendemerits=0
    \medskip\goodbreak\noindent}}

\newcommand{\qedd}{\vrule height4pt width 4pt depth0pt}

\renewcommand{\theta}{\vartheta}
\renewcommand{\phi}{\varphi}

\newcommand{\IC}{\ensuremath{\textsf{IC}_G}\xspace}
\newcommand{\OIC}{\ensuremath{\textsf{IC}^\succ_G}\xspace}
\newcommand{\AIC}{\ensuremath{\textsf{IC}^\axioms_{G,H}}\xspace}

\renewcommand{\exp}[1]{\langle \kern-2pt\langle #1 \rangle \kern-2pt\rangle}
\newcommand{\comp}[1]{\dlb #1 \drb}

\theoremstyle{plain}
\theoremheaderfont{\bf\boldmath}

\newtheorem{theorem}{Theorem}
\newtheorem{definition}{Definition}
\theoremstyle{plain}
\newtheorem{lemma}{Lemma}
\newtheorem{corollary}{Corollary}
\newtheorem{fact}{Fact}
\newtheorem{algorithm}{Algorithm}

\newenvironment{proof}{\vspace{2ex}\noindent \textit{Proof.}~}{}
\newenvironment{claim}{\vspace{2ex}\noindent \textit{Claim.}}{\vspace{2ex}}

\title{The Inverse Method Implements the Automata Approach for Modal
  Satisfiability\thanks{A short version of this report has appeared at the First International Joint Conference on Automated Reasoning, IJCAR 2001. This work has been done while the authors were working at the Research Area for Theoretical Computer Science, RWTH Aachen, Germany.}}

\author{Franz Baader\\
  Theoretical Computer Science, TU Dresden, Germany\\
  \texttt{baader@tcs.inf.tu-dresden.de}
  \and 
  Stephan Tobies\\
  Nokia Research Center, Bochum, Germany\\
  \texttt{stephan.tobies@nokia.com}
}

\begin{document}

\maketitle

\begin{abstract}
  Tableaux-based decision procedures for satisfiability of modal and description
  logics behave quite well in practice, but it is sometimes hard to obtain exact
  worst-case complexity results using these approaches, especially for
  \EXPTIME-complete logics.  In contrast, aut\-o\-mata-based approaches often yield
  algorithms for which optimal worst-case complexity can easily be proved.
  However, the algorithms obtained this way are usually not only worst-case, but
  also best-case exponential: they first construct an automaton that is always
  exponential in the size of the input, and then apply the (polynomial)
  emptiness test to this large automaton. To overcome this problem, one must try
  to construct the automaton ``on-the-fly'' while performing the emptiness test.
 
  In this paper we will show that Voronkov's inverse method for the modal logic
  \K{} can be seen as an on-the-fly realization of the emptiness test done by the
  automata approach for \K{}. The benefits of this result are two-fold. First, it
  shows that Voronkov's implementation of the inverse method, which behaves
  quite well in practice, is an optimized on-the-fly implementation of the
  automata-based satisfiability procedure for \K{}.  Second, it can be used to give
  a simpler proof of the fact that Voronkov's optimizations do not destroy
  completeness of the procedure. We will also show that the inverse method can
  easily be extended to handle global axioms, and that the correspondence to the
  automata approach still holds in this setting. In particular, the inverse
  method yields an \EXPTIME-algorithm for satisfiability in \K{} w.r.t. global
  axioms.
\end{abstract}

\section{Introduction}

Decision procedures for (propositional) modal logics and description logics play
an important r\^{o}le in knowledge representation and verification. When
developing such procedures, one is both interested in their worst-case
complexity and in their behavior in practical applications. From the theoretical
point of view, it is desirable to obtain an algorithm whose worst-case
complexity matches the complexity of the problem. From the practical point of
view it is more important to have an algorithm that is easy to implement and
amenable to optimizations, such that it behaves well on practical instances of
the decision problem.

The most popular approaches for constructing decision procedures for modal
logics are i) semantic tableaux and related methods
\cite{Gore-Tableau-Handbook-1998,BaaderSattler-StudiaLogica-2000}; ii)
translations into classical first-order logics \cite{Schmidt98f,arec:tree00};
and iii) reductions to the emptiness problem for certain (tree)
automata~\cite{VaWo86,LutzSattlerAIML00}.

Whereas highly optimized tableaux and translation approaches behave quite well
in practice \cite{Horrocks-Tableaux-2000,HustadtSchmidt-Tableaux2000}, it is
sometimes hard to obtain exact worst-case complexity results using these
approaches. For example, satisfiability in the basic modal logic \K{} w.r.t.\ 
global axioms is known to be \EXPTIME-complete \cite{Spaan93a}. However, the
``natural'' tableaux algorithm for this problem is a \NEXPTIME-algorithm
\cite{BaaderSattler-StudiaLogica-2000}, and it is rather hard to construct a tableaux
algorithm that runs in deterministic exponential time \cite{donini00:_exptim_alc}.

In contrast, it is folklore that the automata approach yields a very simple proof
that satisfiability in \K{} w.r.t.\ global axioms is in \EXPTIME{}.  However,
the algorithm obtained this way is not only worst-case, but also best-case
exponential: it first constructs an automaton that is always exponential in the
size of the input formulae (its set of states is the powerset of the set of
subformulae of the input formulae), and then applies the (polynomial) emptiness
test to this large automaton. To overcome this problem, one must try to
construct the automaton ``on-the-fly'' while performing the emptiness test.
Whereas this idea has successfully been used for automata that perform model
checking \cite{GPVW95,lncs531*233}, to the best of our knowledge it has not yet
been applied to satisfiability checking.

The original motivation of this work was to compare the automata and the
tableaux approaches, with the ultimate goal of obtaining an approach that
combines the advantages of both, without possessing any of the disadvantages. As
a starting point, we wanted to see whether the tableaux approach could be viewed
as an on-the-fly realization of the emptiness test done by the automata
approach. At first sight, this idea was persuasive since a run of the automaton
constructed by the automata approach (which is a so-called looping automaton
working on infinite trees) looks very much like a run of the tableaux procedure,
and the tableaux procedure does generate sets of formulae on-the-fly.  However,
the polynomial emptiness test for looping automata does \emph{not} try to
construct a run starting with the root of the tree, as done by the tableaux
approach. Instead, it computes inactive states, i.e., states that can never
occur on a successful run of the automaton, and tests whether all initial
states are inactive.  This computation starts ``from the bottom'' by locating
obviously inactive states (i.e., states without successor states), and then
``propagates'' inactiveness along the transition relation. Thus, the
emptiness test works in the opposite direction of the tableaux procedure.  This
observation suggested to consider an approach that inverts the tableaux
approach: this is just the so-called inverse method. Recently, Voronkov
\cite{Voronkov-ToCL-2001} has applied this method to obtain a bottom-up decision
procedure for satisfiability in \K, and has optimized and implemented this
procedure.

In this paper we will show that the inverse method for \K{} can indeed be seen
as an on-the-fly realization of the emptiness test done by the automata approach
for \K{}. The benefits of this result are two-fold. First, it shows that
Voronkov's implementation, which behaves quite well in practice, is an optimized
on-the-fly implementation of the automata-based satisfiability procedure for \K.
Second, it can be used to give a simpler proof of the fact that Voronkov's
optimizations do not destroy completeness of the procedure. We will also show

how the inverse method can be extended to handle global axioms, and that
the correspondence to the automata approach still holds in this setting. In
particular, the inverse method yields an \EXPTIME-algorithm for satisfiability
in \K{} w.r.t.\ global axioms.

\section{Preliminaries}

First, we briefly introduce the modal logic \K{} and some technical definitions
related to \K-formulae, which are used later on to formulate the inverse
calculus and the automata approach for \K.  Then, we define the type of automata
used to decide satisfiability (w.r.t.\ global axioms) in \K. These so-called
looping automata \cite{vardi94:_reason_infin_comput} are a specialization of
B\"uchi tree automata.

\subsubsection*{Modal Formulae}

We assume the reader to be familiar with the basic notions of modal logic.
For a thorough introduction to modal logics, refer to, e.g.,
\cite{blackburn01:_modal_logic}.

\K-formulae are built inductively from a countably infinite set $\mcP = \{ p_1,
p_2, \dots \}$ of propositional atoms using the Boolean connectives $\wedge$,
$\vee$, and $\neg$ and the unary modal operators $\Box$ and $\Diamond$. The
semantics of \K-formulae is define as usual, based on Kripke models $\mcM =
(W,R,V)$ where $W$ is a non-empty set, $R \subseteq W \times W$ is an
accessibility relation, and $V : \mcP \rightarrow 2^W$ is a valuation mapping
propositional atoms to the set of worlds they hold in. The relation $\models$
between models, worlds, and formulae is defined in the usual way. Let $G, H$ be
\K-formulae.

Then $G$ is \emph{satisfiable} iff there exists a Kripke model $\mcM = (W,R,V)$
and a world $w \in W$ with $\mcM, w \models G$. The formula $G$ is
\emph{satisfiable w.r.t.\ the global axiom H} iff there exists a Kripke model
$\mcM = (W,R,V)$ and a world $w \in W$ such $\mcM, w \models G$ and $\mcM, w'
\models H$ for all $w'\in W$.

\K-satisfiability is \textsc{PSpace}-complete~\cite{ladner:1977a},
and \K-satisfiability w.r.t.\ global axioms is \EXPTIME-complete \cite{Spaan93a}. 

A \K-formula is in \emph{negation normal form} (NNF) if $\neg$ occurs only
in front of propositional atoms. Every \K-formula can be transformed (in linear
time) into an equivalent formula in NNF using de Morgan's laws and the duality
of the modal operators. 

For the automata and calculi considered here, sub-formulae of $G$ play an
important role and we will often need operations going from a formula to its
super- or sub-formulae.  As observed in \cite{Voronkov-ToCL-2001}, these
operations become easier when dealing with ``addresses'' of sub-formulae  in $G$

rather than with the sub-formulae themselves.

\begin{definition}[$G$-Paths]
  \label{def:g-paths}
  For a $\K$-formula $G$ in NNF, the \emph{set of $G$-paths} $\Pi_G$ is a set of words
  over the alphabet $\{ {\vee_l}, {\vee_r}, {\wedge_l}, \wedge_r, \Box, \Diamond
  \}$. The set $\Pi_G$ and the sub-formula $G|_\pi$ of $G$ addressed by 
  $\pi \in \Pi_G$ are defined inductively as follows:
  \begin{itemize}
  \item $\epsilon \in \Pi_G$ and $G|_\epsilon = G$
  \item if $\pi \in \Pi_G$ and
    \begin{itemize}
    \item $G|_\pi = F_1 \wedge F_2$ then $\pi {\wedge_l}, \pi {\wedge_r}
      \in \Pi_G$, $G|_{\pi {\wedge_l}} = F_1$, $G|_{\pi {\wedge_r}} =
      F_2$, and $\pi$ is  called \emph{$\wedge$-path}
    \item $G|_\pi = F_1 \vee F_2$ then $\pi {\vee_l}, \pi {\vee_r}
      \in \Pi_G$, $G|_{\pi {\vee_l}} = F_1$, $G|_{\pi {\vee_r}} =
      F_2$, and $\pi$ is called \emph{$\vee$-path}
    \item $G|_\pi = \Box F$ then $\pi \Box \in \Pi_G$, $G|_{\pi \Box}
      = F$ and $\pi$ is  called \emph{$\Box$-path}
    \item $G|_\pi = \Diamond F$ then $\pi \Diamond \in \Pi_G$,
      $G|_{\pi \Diamond} = F$ and $\pi$ is called \emph{$\Diamond$-path}
    \end{itemize}
  \item $\Pi_G$ is the smallest set that satisfies the previous
    conditions.
  \end{itemize}  
\end{definition} 

We use of ${\wedge_*}$ and ${\vee_*}$ as placeholders for ${\wedge_l},
{\wedge_r}$ and ${\vee_l}, {\vee_r}$, respectively. Also, we use $\wedgevee$ and
$\boxdiamond$ as placeholders for $\wedge, \vee$ and $\Box,\Diamond$,
respectively. If $\pi$ is an $\wedge$- or and $\vee$-path then $\pi$ is called
\emph{\wedgevee-path}. If $\pi$ is a $\Box$- or a $\Diamond$-path then $\pi$ is
called \emph{\boxdiamond-path}.

\begin{figure}[t]
  \begin{center}
    \input{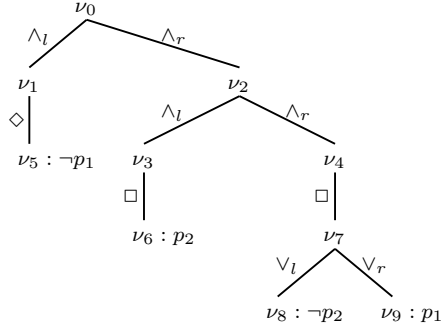}
  \end{center}
  \caption{The set $\Pi_G$ for $G = \Diamond \neg p_1 \wedge (\Box p_2
    \wedge \Box (\neg p_2 \vee p_1))$}
  \label{fig:g-paths}
\end{figure}                     

Figure~\ref{fig:g-paths} shows an example of a \K-formula $G$ and the
corresponding set $\Pi_G$, which can be read off the edge labels. For example,
${\wedge_r}{\wedge_r}$ is a $G$-path and $G|_{{\wedge_r}{\wedge_r}} = \Box (\neg
p_2 \vee p_1)$
 
\subsubsection*{Looping Automata}

For a natural number $n$, let $[n]$ denote the set $\{1, \dots, n\}$. An
\emph{$n$-ary infinite tree over the alphabet $\Sigma$} is a mapping $t : [n]^*
\rightarrow \Sigma$.  An \emph{$n$-ary looping tree automaton} is a tuple $\mfA
= (Q,\Sigma,I, \Delta)$, where $Q$ is a finite set of states, $\Sigma$ is a
finite alphabet, $I \subseteq Q$ is the set of initial states, and $\Delta
\subseteq Q \times \Sigma \times Q^n$ is the transition relation. Sometimes, we
will view $\Delta$ as a function from $Q \times \Sigma$ to $2^{Q^n}$ and write
$\Delta(q,\sigma)$ for the set $\{ \bfq \mid (q,\sigma, \bfq) \in \Delta \}$.

A \emph{run} of $\mfA$ on a tree $t$ is a $n$-ary infinite tree $r$ over $Q$ such that

\[
(r(p),t(p),(r(p1),\dots,r(pn))) \in \Delta 
\]
for every $p \in [n]^*$.  The automaton $\mfA$ \emph{accepts} $t$ iff there is a
run $r$ of $\mfA$ on $t$ such that $r(\epsilon) \in I$.
 The set  $L(\mfA) := \{ t \mid \text{$\mfA$ accepts $t$} \}$ is the \emph{language} accepted by $\mfA$.

Since looping tree automata are special B\"uchi tree automata, emptiness of
their accepted language can effectively be tested using the well-known
(quadratic) emptiness test for B\"uchi automata \cite{VaWo86}. However, for
looping tree automata this algorithm can be specialized into a simpler (linear)
one.  Though this is well-known in the automata theory community, there appears
to be no reference for the result.

Intuitively, the algorithm works by computing inactive states. A state $q \in Q$
is \emph{active} iff there exists a tree $t$ and a run of $\mfA$ on $t$ in which
$q$ occurs; otherwise, $q$ is \emph{inactive}. It is easy to see that a looping
tree automaton accepts at least one tree iff it has an active initial state.
How can the set of inactive states be computed?  Obviously, a state from which
no successor states are reachable is inactive.  Moreover, a state is inactive if
every transition possible from that state involves an inactive state. Thus, one
can start with the set
\[
 Q_0 := \{ q \in Q \mid \forall \sigma \in \Sigma . \Delta(q,\sigma) =
\emptyset \}
\]    
of obviously inactive states, and then propagate inactiveness through the
transition relation.

We formalize this propagation process in a way that allows for an easy
formulation of our main results.

A \emph{derivation} of the emptiness test is a sequence $ Q_0 \vtr Q_1 \vtr
\dots \vtr Q_k$ such that $ Q_i \subseteq Q$ and $ Q_i \vtr Q_{i+1}$ iff $
Q_{i+1} = Q_i \cup \{ q \}$ with
\[
q \in \{ q' \in Q \mid \forall \sigma \in \Sigma . \forall (q_1, \dots, q_n)
\in \Delta(q,\sigma) .  \exists j .  q_j \in  Q_i \} .
\]
We write $Q_0 \vtrstar P$ iff there is a $k \in \N$ and a derivation $Q_0 \vtr
\dots \vtr Q_k$ with $P = Q_k$.  The emptiness test answers ``$L(\mfA) =
\emptyset$'' iff there exists a set of states $P$ such that $ Q_0 \vtrstar P$
and $I \subseteq P$.

Note that $Q \vtr P$ implies $Q \subseteq P$ and that $Q\subseteq Q'$ and $Q
\vtr P$ imply $Q' \vtrstar P$.  Consequently, the \emph{closure} $Q_0^\vtr$ of $Q_0$
under $\vtr$, defined by $ Q_0^\vtr =: \bigcup \{ P \mid Q_0 \vtr P \}$, can
be calculated starting with $Q_0$, and successively adding states $q$ to the
current set $Q_i$ such that $Q_i\vtr Q_i\cup\{q\}$ and $q\not\in Q_i$, until no
more states can be added.  It is easy to see that this closure consists of the
set of inactive states, and thus $L(\mfA) = \emptyset$ iff $I \subseteq
Q_0^\vtr$.  As described until now, this algorithm runs in time polynomial in
the number of states.  By using clever data structures and a propagation
algorithm similar to the one for satisfiability of propositional Horn formulae
\cite{dowling84:_linear_time_testin_horn}, one can in fact obtain a linear
emptiness test for looping tree automata.

\section{Automata, Modal Formulae, and the Inverse Calculus}

We first describe how to decide satisfiability in \K{} using the automata
approach and the inverse method, respectively. Then we show that both
approaches are closely connected.

\subsection{Automata and Modal Formulae}

Given a \K-formula $G$, we define an automaton $\mfA_G$ such
that $L(\mfA_G) = \emptyset$ iff $G$ is not satisfiable. In contrast
to the ``standard''  automata approach, the states of our automaton $\mfA_G$ will
be subsets of $\Pi_G$ rather than sets of subformulae of $G$. 
Using paths instead of subformulae is mostly a matter of notation. We also
require the states to satisfy additional properties (i.e., we do not allow
for arbitrary subsets of $\Pi_G$). This makes the proof of correctness of the
automata approach only slightly more complicated, and it allows us to treat some 
important optimisations of the inverse calculus within our framework.
The next definition introduces these properties.

\begin{definition}[Propositionally expanded, clash]
  Let $G$ be a $\K$-formula in NNF, $\Pi_G$ the set of $G$-paths, and $\Phi
  \subseteq \Pi_G$.  An $\wedge$-path $\pi \in \Phi$ is \emph{propositionally
    expanded in} $\Phi$ iff $\{ \pi {\wedge_l}, \pi {\wedge_r} \} \subseteq
  \Phi$. An $\vee$-path $\pi \in \Phi$ is \emph{propositionally expanded in}
  $\Phi$ iff $\{ \pi {\vee_l} , \pi {\vee_r} \} \cap \Phi \neq \emptyset$.

  The set $\Phi$ is \emph{propositionally expanded} iff every \wedgevee-path $\pi
  \in \Phi$ is propositionally expanded in $\Phi$. We use ``p.e.'' as an
  abbreviation for ``propositionally expanded''.
  
  The set $\Phi'$ is an \emph{expansion} of the set $\Phi$ if $\Phi \subseteq
  \Phi'$, $\Phi'$ is p.e.\ and $\Phi'$ is minimal w.r.t.\ set inclusion with
  these properties.

  For a set $\Phi$, we define the set of its expansions as 
  $\exp \Phi := \{ \Phi' \mid \Phi' \text{ is an expansion of $\Phi$} \}$.

  $\Phi$ contains a \emph{clash} iff there are two paths $\pi_1,\pi_2 \in \Phi$
  such that $G|_{\pi_1} = p$ and $G|_{\pi_2} = \neg p$ for a propositional
  variable $p$. Otherwise, $\Phi$ is called \emph{clash-free}.
\end{definition}  

For a set of paths $\Psi$, the set $\exp \Psi$ can effectively be constructed by
successively adding paths required by the definition of p.e. A formal construction of
the closure can be found in the proof of Lemma~\ref{lem:inverse-calculus-for-propositional-closure}.
Note that $\emptyset$ is p.e., clash-free, and $\exp \emptyset = \{ \emptyset \}$.

\begin{definition}[Formula Automaton]\label{def:automaton}
  For a $\K$-formula $G$ in NNF, we fix an arbitrary enumeration
  $\{ \pi_1, \dots, \pi_n \}$ of the $\Diamond$-paths in $\Pi_G$.

  The $n$-ary looping automaton $\mfA_G$ is defined by $\mfA_G := (Q_G, \Sigma_G,
  \exp{\{ \epsilon \}} , \Delta_G)$, where $Q_G := \Sigma_G := \{ \Phi \subseteq
  \Pi_G \mid \Phi \text{ is p.e.} \}$ and the transition relation $\Delta_G$ is
  defined as follows:
  \begin{itemize}
  \item $\Delta_G$ contains only tuples of the form $(\Phi,\Phi,\dots)$.
  \item If $\Phi$ is clash-free, then we define $\Delta_G(\Phi,\Phi) := \exp
    {\Psi_1} \times  \dots \times \exp{\Psi_n}$, where
    \[
    \Psi_i = \begin{cases} \{ \pi_i \Diamond \} \cup \{ \pi \Box \mid \pi \in
      \Phi \text{ is a $\Box$-path } \} & \text{if
        $\pi_i \in \Phi$ }\\
      \emptyset & \text{else}
    \end{cases}
    \]
  \item If $\Phi$ contains a clash, then $\Delta_G(\Phi,\Phi) = \emptyset$,
    i.e., there is no transition from $\Phi$.
  \end{itemize}
\end{definition}

Note, that this definition implies $\Delta_G(\emptyset,\emptyset) = \{
(\emptyset,\dots,\emptyset) \}$ and only states with a clash have no successor
states.

\begin{theorem}\label{lem:emptiness-and-non-satisfiability}
  For a $\K$-formula $G$, $G$ is satisfiable iff $L(\mfA_G) \neq \emptyset$.
\end{theorem}

This theorem can be proved by showing that 
i) every tree accepted by $\mfA_G$ induces a model of $G$; and 
ii) every model $\mathcal{M}$ of $G$ can be turned into a tree accepted by $\mfA_G$ by 
    a) unraveling $\mathcal{M}$ into a tree model $\mathcal{T}$ for $G$; 
    b) labeling every world of $\mathcal{T}$ with a suitable p.e. set depending on
       the formulae that hold in this world; and 
    c) padding ``holes'' in $\mathcal{T}$ with $\emptyset$.

\begin{proof}
  Let $\{\pi_1, \dots, \pi_n\}$ be an enumeration of the $\Diamond$-paths in
  $\Pi_G$.
  
  For the \emph{if}-direction let $L(\mfA_G) \neq \emptyset$, $t,r : [n]^*
  \rightarrow \{ \Phi \subseteq \Pi_G \mid \Phi \text{ is p.e.} \}$ a tree that
  is accepted by $\mfA_G$ and a corresponding run of $\mfA_G$. By construction
  of $\mfA_G$, $t(w) = r(w)$ for every $w \in [n]^*$.  We construct a Kripke
  model $\mcM = (W,R,V)$ from $t$ by setting
  \begin{align*}
    W & = \{ w \in [n]^* \mid t(w) \neq \emptyset \}\\
    R & = \{ (w,wi) \in W \times W \mid i \in [n] \}\\
    V & = \lambda P . \{ p \in W \mid \exists \pi \in t(w) . G|_\pi = P \} \quad
    \text{ for all propositional atoms $P$ }
  \end{align*}
  
  \begin{claim}
    For all $w \in W$, if $\pi \in t(w)$ then $\mcM,w \models G|_\pi$.
  \end{claim}
  
  \noindent\textit{Proof of the claim.} The claim is proved by induction on the
  structure of \K-formulae.  Let $w \in W$ be a world and $\pi \in \Pi_G$ be a
  path such that $\pi \in t(w)$.
  \begin{itemize}
  \item if $G|_\pi = P$ is a propositional atom and $w \in W$, then $w \in V(P)$
    and hence $\mcM,w \models G|_\pi$.
  \item if $G|_\pi = \neg P$ is a negated propositional atom, then, since $t(w)$
    is clash free, there is no $\pi' \in t(w)$ such that $G|_{\pi'} = P$. Thus,
    $w \not \in V(P)$ and hence $\mcM,w \models \neg P$.
  \item if $G|_\pi = F_1 \wedge F_2$ then $\pi$ is an $\wedge$-path, and since
    $t(w)$ is p.e., $\{ \pi {\wedge_l}, \pi {\wedge_r}\} \subseteq t(w)$. By
    induction, $\mcM, w \models G|_{\pi{\wedge_*}}$ and hence $\mcM,w \models
    G|_\pi$.
  \item if $G|_\pi = F_1 \vee F_2$ then $\pi$ is an $\vee$-path, and since
    $t(w)$ is p.e., $\{ \pi {\vee_l}, \pi {\vee_r}\} \cap t(w) \neq \emptyset$.
    By induction, $\mcM, w \models G|_{\pi{\vee_l}}$ or $\mcM, w \models
    G|_{\pi{\vee_r}}$ and hence $\mcM,w \models G|_\pi$.
  \item if $G|_\pi = \Diamond F$ then $\pi$ is a $\Diamond$-path and, w.o.l.g.,
    assume $\pi = \pi_i$. Since $\pi_i \in r(w)$, $\pi_i \Diamond \in r(wi) =
    t(wi)$ holds and hence $wi \in W$ and $(w,wi) \in R$. By induction, we have
    that $\mcM, wi \models G|_{\pi_i \Diamond}$ and hence $\mcM, w \models
    G|_{\pi_i}$.
  \item if $G|_\pi = \Box F$ and $(w,w') \in R$ then $w' = wi$ for some $i \in
    [n]$ and $t(wi) \neq \emptyset$ holds and by construction of $\mfA_G$, this
    implies $\pi \Box \in r(wi) = t(wi)$. By induction, this implies $\mcM, wi
    \models G|_{\pi \Box}$ and since $wi = w'$ and $w'$ has been chosen
    arbitrarily, $\mcM, w \models G|_\pi$.
  \end{itemize}
  
  This finishes the proof of the claim. Since $t(\epsilon) = r(\epsilon) \in
  \exp{ \{ \epsilon \} }$ and hence $\epsilon \in t(\epsilon)$, $\mcM, \epsilon
  \models G|_\epsilon$ and $G = G|_\epsilon$ is satisfiable.
  
  For the \emph{only if}-direction, we first show an auxiliary claim: for a set
  $\Psi \subseteq \Pi_G$ we define $\mcM, w \models \Psi$ iff $\mcM, w \models
  G|_\pi$ for every $\pi \in \Psi$.

  \begin{claim}
    If $\Psi \subseteq \Pi_G$ and $w \in W$ such that $\mcM, w \models \Psi$, then there
    is a $\Phi \in \exp \Psi$ such that $\mcM, w \models \Phi$.
  \end{claim}
  
  \noindent \textit{Proof of the claim.} Let $\Psi \subseteq \Pi_G$ and $w \in
  W$ such that $\mcM, w \models \Psi$.  We will show how to construct an
  expansion of $\Psi$ with the desired property.  If $\Psi$ is already p.e.,
  then $\Psi \in \exp \Psi$ and we are done. If $\Psi$ is not p.e.\ then let
  $\pi \in \Psi$ be a \wedgevee-path that is not p.e.\ in $\Psi$.
  \begin{itemize}
  \item If $\pi$ is a $\wedge$-path then $G|_\pi = F_1 \wedge F_2$ and since
    $\mcM, w \models G|_\pi$, also $\mcM, w \models F_1 = G|_{\pi \wedge_l}$ and
    $\mcM, w \models F_2 = G|_{\pi \wedge_r}$. Hence $\mcM, w \models \Psi \cup
    \{ \pi{\wedge_l}, \pi{\wedge_r} \}$ and $\Psi' = \Psi \cup \{ \pi{\wedge_l},
    \pi{\wedge_r}\}$ is a set with $\mcM, w \models \Psi'$ that is``one step
    closer'' to being p.e.\ than $\Psi$.
  \item If $\pi$ is a $\vee$-path then $G|_\pi = F_1 \vee F_2$ and since $\mcM,
    w \models G|_\pi$, also $\mcM, w \models F_1 = G|_{\pi \vee_l}$ or $\mcM, w
    \models F_2 = G|_{\pi \vee_r}$. Hence $\mcM, w \models \Psi \cup \{
    \pi{\vee_l}\}$ or $\mcM, w \models \Psi \cup \{ \pi{\vee_r}\}$ and hence can
    obtain a set $\Psi'$ with $\mcM, w \models \Psi'$ that is again ``one step
    close'' to being p.e. than $\Psi$.
  \end{itemize}
  
  Restarting this process with $\Psi = \Psi'$ eventually yields an expansion
  $\Phi$ of the initial set $\Psi$ with $\mcM,w \models \Phi$, which proves the
  claim.
  
  Let $\mcM = (W,R,V)$ be a model for $G$ with $w \in W$ such that $\mcM, w
  \models G$. From $\mcM$ we construct a tree that is accepted by $\mfA_G$.
  Using this claim, we inductively define a tree $t$ accepted by $\mfA_G$. To
  this purpose, we also inductively define a function $f : [n]^* \rightarrow W$
  such that, if $\mcM, f(p) \models t(p)$ for all $p$.
  
  We start by setting $f(\epsilon) = w$ for a $w \in W$ with $\mcM, w \models
  G$.  and $t(\epsilon) = \Phi$ for a $\Phi \in \exp { \{ \epsilon \} }$ such
  that $\mcM, w \models \Phi$. From the claim we have that such a set $\Phi$
  exists because $\mcM, w \models G = G|_\epsilon$.
  
  If $f(p)$ and $t(p)$ are already defined, then, for $i \in [n]$, we define
  $f(pi)$ and $t(pi)$ as follows:
  \begin{itemize}
  \item if $\pi_i \in t(p)$ then $\mcM, f(p) \models G|_{\pi_i}$ and hence there
    is a $w' \in W$ such that $(f(p),w') \in R$ and $\mcM, w' \models G|_{\pi_i
      \Diamond}$. If $\pi \in t(p)$ is a $\Box$-path, then also $\mcM, w'
    \models G|_{\pi\Box}$ holds. Hence $\mcM, w' \models \{ \pi_i \Diamond \}
    \cup \{ \pi \Box \mid \pi \in t(p) \text{ is a $\Box$-path } \}$.  We set
    $f(pi) = w'$ and $t(pi) = \Phi$ for a $\Phi \in \exp {\{ \pi_i \Diamond \}
      \cup \{ \pi \Box \mid \pi \in t(p) \text{ is a $\Box$-path } \}}$ with
    $\mcM, w' \models \Phi$, which exist by the claim.
  \item if $\pi_i \not \in t(p)$, then we set $f(pi) = w$ for an arbitrary $w
    \in W$ and $t(pi) = \emptyset$.
  \end{itemize}
  In both cases, we have define $f(pi)$ and $t(pi)$ such that $\mcM, f(pi)
  \models t(pi)$. It is easy to see that $t$ is accepted by $\mfA_G$ with the
  run $r = t$. Hence $L(\mfA_G) \neq \emptyset$ which is what we needed to show.
  \qed
\end{proof}

Together with the emptiness test for looping tree automata,
Theorem~\ref{lem:emptiness-and-non-satisfiability} yields a decision procedure
for $\K$-satisfiability. To test a \K-formula $G$ for unsatisfiability,
construct $\mfA_G$ and test whether $L(\mfA_G)=\emptyset$ holds using the
emptiness test for looping tree automata: 
$L(\mfA_G)=\emptyset$ iff $\exp{ \{ \epsilon \}}\subseteq Q_0^\vtr$,
where $Q_0 \subseteq Q_G$ is the set of states containing a clash. 

The following is a derivation of a superset of $\exp {\{ \epsilon \}}$ from $Q_0$ 
for the example formula from Figure~\ref{fig:g-paths}:

$$
      Q_0  = \{ \underbrace{\{ \nu_5,\nu_6,\nu_7, \nu_8 \},\ \{
        \nu_5,\nu_6,\nu_7, \nu_9 \}}_{=\ \exp { \nu_5, \nu_6, \nu_7 }},
      \dots \} 
       \vtr Q_0 \cup \underbrace{ \{ \{ \nu_0, \nu_1, \nu_2, \nu_3, \nu_4 \}\}}_{=\
        \exp { \{\epsilon \} }}                       
$$

\subsection{The Inverse Calculus}

In the following, we introduce the inverse calculus for \K. We stay close to
the notation and terminology used in \cite{Voronkov-ToCL-2001}.  

A \emph{sequent} is a subset of $\Pi_G$. Sequents will be denoted by capital greek
letters. The union of two sequents $\Gamma$ and $\Lambda$ is denote by $ \Gamma,
\Lambda$.  If $\Gamma$ is a sequent and $\pi \in \Pi_G$ then we denote $\Gamma
\cup \{ \pi \}$ by $\Gamma, \pi$. 

If $\Gamma$ is a sequent that contains only $\Box$-paths then we write $\Gamma
\Box$ to denote the sequent $\{ \pi \Box \mid \pi \in \Gamma \}$.  Since states
of $\mfA_G$ are also subsets of $\Pi_G$ and hence sequents, we will later on
use the same notational conventions for states as for sequents.

\begin{definition}[The inverse path calculus]
  Let $G$ be a formula in NNF and $\Pi_G$ the set of paths of $G$.

  \emph{Axioms} of the inverse calculus are all sequents $\{ \pi_1, \pi_2 \}$ such that
  $G|_{\pi_1} = p$ and $G|_{\pi_2} = \neg p$ for some propositional variable
  $p$. The \emph{rules} of the inverse calculus are given in Figure \ref{fig:inv-calc},
  where all paths occurring in a sequent are $G$-paths and, for every
  $\Diamond^+$ inference, $\pi$ is a $\Diamond$-path.  We refer to this calculus
  by \IC.\footnote{$G$ appears in the subscript because the calculus is highly
    dependent of the input formula $G$: only $G$-paths can be generated by \IC.}

  We define $\mcS_0 := \{ \Gamma \mid \Gamma \text{ is an axiom } \}$. A
  \emph{derivation} of \IC is a sequence of sets of sequents $\mcS_0 \vdash \dots
  \vdash \mcS_m$ where $\mcS_i \vdash \mcS_{i+1}$ iff $\mcS_{i+1} = \mcS_i \cup
  \{ \Gamma \}$ such that there exists sequents $\Gamma_1, \dots \Gamma_k \in \mcS_i$
  and $\inference{\Gamma_1 & \dots & \Gamma_k}{\Gamma}$ is an inference.
\end{definition}

We write $\mcS_0 \vdstar \mcS$ iff there is a derivation
$\mcS_0 \vdash \dots \vdash \mcS_m$ with $\mcS = \mcS_m$.  The \emph{closure}
$\mcS_0^\vdash$ of $\mcS_0$ under $\vdash$ is defined by 
$\mcS_0^\vdash = \bigcup \{ \mcS \mid \mcS_0 \vdstar \mcS \}$. 
Again, the closure can effectively be computed by starting with $\mcS_0$ and
then adding sequents that can be obtained by an inference until no more new
sequents can be added. 

  \begin{figure}[t]
    \begin{center}
      \begin{tabular}{c@{\ \ \ }c@{\ \ \ }c}
        \inference[$(\vee)$] {\Gamma_l, \pi \vee_l & \Gamma_r,\pi
          \vee_r}{\Gamma_l, \Gamma_r, \pi} 
         & \inference[$(\wedge_l)$]{ \Gamma, \pi \wedge_l}{\Gamma, \pi} &
        \inference[$(\wedge_r)$]{ \Gamma, \pi \wedge_r}{\Gamma, \pi} \\[3ex]
        \inference[$(\Diamond)$]{\Gamma \Box, \pi \Diamond}{\Gamma, \pi} &
        \inference[$(\Diamond^+)$]{\Gamma \Box}{\Gamma, \pi}
      \end{tabular}
    \end{center}
    \vspace{-1ex}
    \caption{Inference rules of \IC}
    \vspace{-2ex}
    \label{fig:inv-calc}
  \end{figure}

As shown in \cite{Voronkov-ToCL-2001}, the computation of the closure yields
a decision procedure for \K-satisfiability:

\begin{fact}\label{fact:inverse-calculus-decides-k-satisfiability}
  $G$ is unsatisfiable iff $\{ \epsilon \} \in \mcS_0^\vdash$.
\end{fact}

Figure~\ref{fig:example-inference} shows the inferences of \IC that lead to
$\nu_0 = \epsilon$ for the example formula from Figure~\ref{fig:g-paths}.

\subsection{Connecting the Two Approaches}

The results shown in this subsection imply that \IC can be viewed as an on-the-fly
implementation of the emptiness for $\mfA_G$. 

In addition to generating states on-the-fly, states are also represented in a
 compact manner: one sequent generated by \IC represents several states
of $\mfA_G$.

\begin{definition}
  For the formula automaton $\mfA_G$ with states $Q_G$ and a sequent $\Gamma
  \subseteq \Pi_G$ we define $\comp \Gamma := \{ \Phi \in Q_G \mid
  \Gamma \subseteq \Phi \}$, and for a set $\mcS$ of sequents we define $ \comp \mcS
   := \bigcup_{\Gamma \in \mcS} \comp \Gamma$.
\end{definition}

The following theorem, which is one of the main contributions of this paper,
establishes the correspondence between the emptiness test and \IC. 

\begin{theorem}[\IC and the emptiness test mutually simulate each other]%
\label{theo:inv-calc-to-emptiness-test}\label{theo:emptiness-test-to-inv-calc}%
Let 
$Q_0$, $\mcS_0$, $\vtr$, and $\vdash$ be defined as above.
  \begin{enumerate}
  \item   
    Let $Q$ be a set of states such that $Q_0 \vtrstar Q$. Then there exists a set
    of sequents $\mcS$ with 
    $\mcS_0 \vdstar \mcS \text{ and } Q \subseteq \comp \mcS$.
  \item  Let $\mcS$ be a set of sequents
    such that $\mcS_0 \vdstar \mcS$. Then there exists a set of states $Q
    \subseteq Q_G$ with
    $Q_0 \vtrstar Q \text{ and } \comp \mcS \subseteq Q$.
  \end{enumerate}
\end{theorem}

The first part of the theorem shows that \IC can simulate each computation of the
emptiness test for $\mfA_G$. The set of states represented by the set of sequents
computed by \IC may be larger than the one computed by a particular derivation of
the emptiness test. However, the second part of the theorem implies that all these
states are in fact inactive since a possibly larger set of states can also be
computed by a derivation of the emptiness test.

In particular, the theorem implies that \IC can be used to calculate a compact
representation of $Q_0^\vtr$.  This is an on-the-fly computation since $\mfA_G$
is never constructed explicitly.

\begin{corollary}\label{cor:closures-equal}
  $Q_0^\vtr  = \comp {\mcS_0^\vdash}$.
\end{corollary}

\begin{figure}[t]
  \begin{center}
    $
    \inference[$(\vee)$]{{\wedge_l}\Diamond,\ {\wedge_r}{\wedge_r}\Box{\vee_r} &
      | &
      {\wedge_r}{\wedge_l}\Box,\  {\wedge_r}{\wedge_r}\Box{\vee_l}}
    {\inference[$(\Diamond)$]{{\wedge_l}\Diamond,\ {\wedge_r}{\wedge_l}\Box, \ ,
        {\wedge_r}{\wedge_r}\Box} 
      {\inference[$(\wedge_r)$]{{\wedge_l},\ {\wedge_r}{\wedge_l},\ 
          {\wedge_r}{\wedge_r}}
        {\inference[$(\wedge_l)$]{{\wedge_l}, \ {\wedge_r},\
            {\wedge_r}{\wedge_l}}
          {\inference[$(\wedge_r)$]{{\wedge_l},\ {\wedge_r}}
            {\inference[$(\wedge_l)$]{\epsilon,\ {\wedge_l}}
              {\epsilon}}}}}}
    $
  \end{center}
  \vspace{-1ex}
  \caption{An example of inferences in \IC}
  \vspace{-1ex}
  \label{fig:example-inference}
\end{figure}

\begin{proof} 
  If $\Phi \in Q_0^\vtr$ then there exists a set of states $Q$ such that $Q_0
  \vtrstar Q$ and $\Phi \in Q$. By
  Theorem~\ref{theo:emptiness-test-to-inv-calc}.1, there exists a set of sequents
  $\mcS$ with $\mcS_0 \vdstar\mcS$ and $Q \subseteq \comp \mcS$. Hence $\Phi
  \in \comp {\mcS_0^\vdash}$. For the converse direction, if $\Phi \in \comp
  {\mcS_0^\vdash}$ then there exists a set of sequents $\mcS$ with $\mcS_0
  \vdstar \mcS$ and $\Phi \in \comp \mcS$. By
  Theorem~\ref{theo:inv-calc-to-emptiness-test}.2, there exists a set of states
  $Q$ with $Q_0 \vtrstar Q$ and $\comp \mcS \subseteq Q$ and hence $\Phi \in
  Q_0^\vtr$. \qed
\end{proof}

The proof of the second part of Theorem~\ref{theo:inv-calc-to-emptiness-test} is
the easier one. It is a consequence of the next three lemmata.  First, observe
that the two calculi have the same starting points.

\begin{lemma}\label{lem:dead-states-equal-axioms}
  If $\mcS_0$ is the set of axioms of \IC, and $ Q_0$ is the set of states of
  $\mfA_G$ that have no successor states, then $\comp {\mcS_0}=  Q_0$.
\end{lemma}

\begin{proof} 
  The set $\mcS_0$ is the set of all axioms i.e., the set of all clashes.  Hence
  $\comp{ \mcS_0} = \{ \Phi \mid \Phi \text{ contains a clash} \} = Q_0$. \qed
\end{proof}

Second, since states are assumed to be p.e.,
propositional inferences of \IC do not change the set of states
represented by the sequents.

\begin{lemma}\label{lem:propositional-inferences-for-free}
  Let $\mcS \vdash \mcT$ be a derivation of \IC that employs a
  $\wedge_l$-, $\wedge_r$-, or a $\vee$-inference. Then $\comp \mcS = \comp \mcT$.
\end{lemma}

\begin{proof} 
  Since $\mcS \subseteq \mcT$, $\comp \mcS \subseteq \comp \mcT$ holds
  immediately. To show $\comp \mcT \subseteq \comp \mcS$, we distinguish the
  different inferences used to obtain $\mcT$ from $\mcS$:
  \begin{itemize}
  \item If the employed inference is $\inference[$(\wedge_*)$]{\Gamma, \pi
      \wedge_*}{\Gamma,\pi}$ and $\mcT = \mcS \cup \{ \Gamma, \pi \}$ with
    $\Gamma, \pi \wedge_* \in \mcS$. Then $\comp \mcT = \comp \mcS \cup \comp
    {\Gamma,\pi}$. Let $\Phi \in \comp {\Gamma,\pi}$. $\Phi$ is p.e.\ and hence
    $\pi \in \Phi$ implies $\pi\wedge_* \in \Phi$. Thus, $\Gamma,\pi\wedge_*
    \subseteq \Phi$ and $\Phi \in \comp{\Gamma,\pi\wedge_*} \subseteq \comp
    \mcS$.
  \item Assume that the employed inference is $\inference[$(\vee)$]{\Gamma_l,
      \pi \vee_l & \Gamma_r, \vee_r}{\Gamma_l, \Gamma_r,\pi}$ and $\mcT = \mcS
    \cup \{ \Gamma_l, \Gamma_r, \pi \}$ with $\Gamma_l, \pi \vee_l \in \mcS$,
    $\Gamma_r, \vee_r \in \mcS$. Then $\comp \mcT = \comp \mcS \cup \comp
    {\Gamma_l,\Gamma_r,\pi}$. Let $\Phi \in \comp {\Gamma_l,\Gamma_r,\pi}$.
    $\Phi$ is p.e.\ and hence, w.o.l.g., $\pi \vee_l \in \Phi$. Thus, $\Gamma_l,
    \pi \vee_l \subseteq \Phi$ and $\Phi \in \comp { \Gamma_l, \pi \vee_l }
    \subseteq \comp \mcS$. \qed
  \end{itemize}
\end{proof}

Third, modal inferences of \IC can be simulated by derivations of the emptiness test.

\begin{lemma}\label{lem:modal-inference-simulated-by-emptiness-test}
  Let $\mcS \vdash \mcT$ be derivation of \IC that employs a
  $\Diamond$- or $\Diamond^+$-inference. If $Q$ is a set of states with
  $\comp \mcS \cup  Q_0 \subseteq  Q$ then there exists a set of states $P$ with
  $Q \vtrstar P$ and $\comp \mcT \subseteq  P$.
\end{lemma}

\begin{proof} 
  We only consider the $\Diamond$-inference, the case of a
  $\Diamond^+$-inference is analogous.  If $\mcS \vdash \mcT$ by an application of
  a $\Diamond$-inference, then $\mcT = \mcS \cup \{ \Gamma,\pi \}$ where
  $\Gamma$ consists only of $\Box$-paths, $\pi$ is a $\Diamond$-path (w.o.l.g.,
  we assume $\pi = \pi_i$, the $i$-th path in the enumeration of
  $\Diamond$-paths in $\Pi_G$), $\Gamma \Box, \pi_i \Diamond \in \mcS$ and
  $\inference[$(\Diamond)$]{\Gamma \Box, \pi_i \Diamond}{\Gamma,
    \pi_i}$. Also, $\comp \mcT = \comp \mcS \cup \comp { \Gamma, \pi_i }$ holds.

  \begin{claim}
    Let $\Phi \in \comp { \Gamma,\pi_i }$ and $R$ a set of states with $\comp {
      \Gamma \Box, \pi_i \Diamond} \cup Q_0 \subseteq R$. Then there exists a
    derivation $ R \vtrstar R'$ with $\Phi \in R'$ and $\comp { \Gamma \Box,
      (\pi_i \Diamond)} \cup Q_0 \subseteq R'$
  \end{claim}
  
  \noindent \textit{Proof of the Claim.}  If $\Phi$ contains a clash then $\Phi
  \in Q_0 \subseteq R$ and nothing has to be done.  If $\Phi$ does not contain a
  clash, then $\Delta_G(\Phi,\Phi) = \exp {\Psi_i } \times \dots \times \exp
  {\Psi_n}$ where the $\Psi_i$ are defined as in Definition \ref{def:automaton}
  and especially, since $\pi_i \in \Phi$,
  \[
  \exp{\Psi_i} = \exp{ \underbrace{\{ \pi_i \Diamond \} \cup \{ \pi \Box \mid
      \pi \in \Phi \text{ is a $\Box$-path } \}}_{\supseteq \Gamma \Box, \pi_i
      \Diamond}} \subseteq \comp {\Gamma \Box, \pi_i \Diamond} \subseteq R
  \]
  Since all states in $\exp {\Psi_i}$ have been marked inactive, the emptiness
  test can also mark $\Phi$ inactive and derive $ R \vtr R \cup \{ \Phi \} =
  R'$, which proves the claim.
  
  Using this claim, we prove the lemma as follows. Let $\Phi_i, \dots \Phi_k$ be
  an enumeration of $\comp {\Gamma, \pi_i}$. The set $P_0 = Q$ satisfies the
  requirements of the claim for $ R$. Thus, we repeatedly use the claim and
  chain the derivations to obtain a derivation $ Q = P_0 \vtr P_1 \vtr \dots
  \vtr P_k = P$ such that $\Phi_i \in P_i$. Since the sets grow monotonically, in
  the end $\comp {\Gamma,\pi} \subseteq P$ holds, which implies $\comp \mcT
  \subseteq P$. \qed
\end{proof}

Given these lemmata, proving
Theorem~\ref{theo:inv-calc-to-emptiness-test}.2 is quite simple.

\paragraph{
Proof of Theorem~\ref{theo:inv-calc-to-emptiness-test}.2.}
The proof is by induction on the length $m$ of the derivation $\mcS_0 \vdash \mcS_1
\dots \vdash \mcS_m = \mcS$ of \IC . The base case $m=0$ is
Lemma~\ref{lem:dead-states-equal-axioms}. For the induction step, $\mcS_{i+1}$
is either inferred from $\mcS_i$ using a propositional inference, which is dealt
with by Lemma~\ref{lem:propositional-inferences-for-free}, or by a modal
inference, which is dealt with by
Lemma~\ref{lem:modal-inference-simulated-by-emptiness-test}.
Lemma~\ref{lem:modal-inference-simulated-by-emptiness-test} is applicable
since, for  every set of states $Q$ with  $Q_0 \vtrstar Q$, 
$Q_0 \subseteq Q$. \qed

\medskip

Proving the first part of Theorem~\ref{theo:inv-calc-to-emptiness-test} is more involed
because of the calculation of the propositional expansions implicit in the definition
of $\mfA_G$.

\begin{lemma}\label{lem:inverse-calculus-for-propositional-closure}
  Let $\Phi \subseteq \Pi_G$ be a set of paths and $\mcS$ a set of sequents such
  that $\exp \Phi \subseteq \comp \mcS$. Then there exists a set of sequents
  $\mcT$ with $\mcS \vdstar \mcT$ such that there exists a sequent $\Lambda \in \mcT$
  with $\Lambda \subseteq \Phi$.
\end{lemma}

\begin{proof}If $\Phi$ is p.e., then this is immediate, as in this case $\exp {\Phi} = \{
  \Phi \} \subseteq \comp \mcS$.
  
  If $\Phi$ is not p.e., then let $\select$ be an arbitrary \emph{selection
    function}, i.e., a function that maps every set $\Psi$ that is not p.e.\ to
  a \wedgevee-path $\pi\in \Psi$ that is not p.e.\ in $\Psi$. Let $\Tree_\Phi$
  be the following, inductively defined tree:
  \begin{itemize}
  \item The root of $\Tree_\Phi$ is $\Phi$.
  \item If a node $\Psi$ of $\Tree_\Phi$ is not p.e., then
    \begin{itemize}
    \item if $\select(\Psi) = \pi$ is an $\wedge$-path, then $\Psi$ has the
      successor node $\Psi, \pi \wedge_l, \pi\wedge_r$ and $\Psi$ is called
      an $\wedge$-node.
    \item if $\select(\Psi) = \pi$ is an $\vee$-path, then $\Psi$ has the
      successor nodes $\Psi, \pi\vee_l$ and $\Psi, \pi\vee_l$ and $\Psi$ is called
      an $\vee$-node.
    \end{itemize}
  \item If a node $\Psi$ of $\Tree_\Phi$ is p.e., then it is a leaf of the tree.
  \end{itemize}
  Obviously, the construction is such that the set of leaves of $\Tree_\Phi$ is
  $\exp \Phi$.  
  
  Let $\Upsilon_1, \dots \Upsilon_\ell$ be a post-order traversal of this tree,
  so the sons of a node occur before the node itself and $\Upsilon_\ell = \Phi$.

  Along this traversal we will construct a derivation $\mcS = \mcT_0 \vdstar
  \dots \vdstar \mcT_\ell = \mcT$ such that, for every $1\leq i \leq j \leq
  \ell$, $\mcT_j$ contains a sequent $\Lambda_i$ with $\Lambda_i \subseteq
  \Upsilon_i$.  Since the sets $\mcT_j$ grow monotonically, it suffices to show
  that, for every $1 \leq i \leq \ell$, $\mcT_i$ contains a sequent $\Lambda_i$
  with $\Lambda_i \subseteq \Upsilon_i$.
  
  Whenever $\Upsilon_i$ is a leaf of $\Tree_\Phi$, then $\Upsilon_i \in
  \exp{\Phi} \subseteq \comp \mcS$. Hence there is already a sequent $\Lambda_i
  \in \mcT_0$ with $\Lambda_i \subseteq \Upsilon_i$ and no derivation step is
  necessary. Particularly, in a post-order traversal, $\Upsilon_1$ is a leaf.

  We now assume that the derivation has been constructed up to $\mcT_i$.

  \begin{itemize}
  \item If $\Upsilon_{i+1}$ is a leaf of $\Tree_\Phi$, then nothing has to be
    done as there exists a $\Lambda_{i+1} \in \mcT_0 \subseteq \mcT_i$ with
    $\Lambda_{i+1} \subseteq \Upsilon_{i+1}$
  \item If $\Upsilon_{i+1}$ is an $\wedge$-node with selected $\wedge$-path $\pi
    \in \Upsilon_{i+1}$. Then, the successor of $\Upsilon_{i+1}$ in $\Tree_\Phi$
    is $\Upsilon_{i+1} \pi\wedge_l, \pi\wedge_r$ and appears before
    $\Upsilon_{i+1}$ in the traversal. By construction there exists a sequent
    $\Lambda \in \mcT_i$ with $\Lambda \subseteq \Upsilon_{i+1},\pi \wedge_l,
    \pi \wedge_r$. If $\Lambda \cap \{ \pi \wedge_l, \pi \wedge_r \} =
    \emptyset$ then we are done because then also $\Lambda \subseteq
    \Upsilon_{i+1}$. If one or both of $\pi \wedge_l, \pi \wedge_r$ occur in
    $\Lambda$, then
    \begin{itemize}
    \item if $\Lambda = \Gamma, \pi \wedge_l$ for some $\Gamma$ with $\pi
      \wedge_r \not \in \Gamma$ then this implies that the inference
      \begin{equation}\label{eq:and-inference}
        \inference[$(\wedge_l)$]{\Gamma, \pi \wedge_l}{\Gamma, \pi}
      \end{equation}
      can be used to derive $\mcT_i \vdash \mcT_i \cup \{ \Gamma, \pi \} =
      \mcT_{i+1}$ and $\Gamma, \pi \subseteq \Upsilon_{i+1}$ holds.
    \item the case $\Lambda = \Gamma, \pi \wedge_r$ for some $\Gamma$ with $\pi
      \wedge_l \not \in \Gamma$ if analogous.
    \item if $\Lambda = \Gamma, \pi {\wedge_l}, \pi {\wedge_r}$ for some
      $\Gamma$ with $\{\pi {\wedge_l}, \pi {\wedge_r}\} \cap \Gamma = \emptyset$
      then the inferences
      \begin{equation}\label{eq:and-inference-2}
        \inference[$(\wedge_l)$]{ \Gamma, \pi \wedge_l, \pi
          \wedge_r}{ \inference[$(\wedge_r)$]{\Gamma, \pi, \pi \wedge_r}{\Gamma, \pi, \pi}}
      \end{equation}
      can be used in the derivation $\mcT_i \vdash \mcT_i \cup \{ \Gamma, \pi,
      \pi \wedge_r \} \vdash \mcT_i \cup \{\Gamma, \pi, \pi \wedge_r \} \cup \{
      \Gamma, \pi \} = \mcT_{i+1} $ and by construction $\Gamma, \pi \subseteq
      \Upsilon_{i+1}$ holds.
    \end{itemize}
  \item If $\Upsilon_{i+1}$ is an $\vee$-node with selected $\vee$-path $\pi \in
    \Upsilon_{i+1}$. Then, the successors of $\Upsilon_{i+1}$ in $\Tree_\Phi$
    are $\Upsilon_{i+1}, \pi\vee_l$ and $\Upsilon_{i+1}, \pi\vee_r$, and by
    construction there exist sequences $\Lambda_l, \Lambda_r \in \mcT_i$ with
    $\Lambda_* \subseteq \Upsilon_{i+1}, \pi\vee_*$.
  
    If $\pi \vee_l \not \in \Lambda_l$ or $\pi \vee_r \not \in \Lambda_r$, then
    $\Lambda_l \subseteq \Upsilon_{i+1} $ or $\Lambda_r \subseteq
    \Upsilon_{i+1}$ holds and hence already $\mcT_i$ contains a sequent
    $\Lambda$ with $\Lambda \subseteq \Upsilon_{i+1}$.
  
    If $\Lambda_l = \Gamma_l, \pi \vee_l$ and $\Lambda_r = \Gamma_r, \pi \vee_r$
    with $\pi {\vee_*} \not \in \Gamma_*$ then \IC can use the inference
    \begin{equation}\label{eq:or-inference}
      \inference[$(\vee)$]{\Gamma_l, \pi \vee_l & \Gamma_r, \pi
        \vee_r}{\Gamma_l, \Gamma_r, \pi}
    \end{equation}
    to derive $\mcT_i \vdash \mcT_i \cup \{ \Gamma_l, \Gamma_r, \pi \} =
    \mcT_{i+1}$, and and $\Gamma_l, \Gamma_r, \pi \subseteq \Upsilon_{i+1}$
    holds as follows: assume there is a $\pi' \in \Gamma_l, \Gamma_r, \pi$ with
    $\pi' \not \in \Upsilon_{i+1}$.  Since $\pi \in \Upsilon_{i+1}$, w.o.l.g.,
    $\pi' \in \Gamma_l$.  But then also $\Gamma_l \not \subseteq \Upsilon_{i+1},
    \pi \vee_l $ would hold, since $\pi' \neq \pi {\vee_l}$ because $\pi
    {\vee_l} \not \in \Gamma_l$.
  \end{itemize}
  
  Proceeding in this manner, starting from $\mcT_0 = \mcS$, we can construct a
  derivation that yields a set $\mcT = \mcT_k$ of states containing a sequent
  $\Lambda$ such that $\Lambda \subseteq \Upsilon_\ell = \Phi$. \qed
\end{proof}

\paragraph{\textit{Proof of Theorem~\ref{theo:emptiness-test-to-inv-calc}.1.}}
We show this by induction on the number $k$ of steps in the derivation $Q_0 \vtr
\dots \vtr Q_k = Q$. Again, Lemma~\ref{lem:dead-states-equal-axioms} yields the
base case.

For the induction step, let $Q_0 \vtr \dots \vtr Q_i \vtr Q_{i+1} = Q_i \cup \{
\Phi \}$ be a derivation of the emptiness test and $\mcS_i$ a set of sequents
such that $\mcS \vdstar \mcS_i$ and $ Q_i \subseteq \comp{ \mcS_i}$.  Such a set
exists by the induction hypothesis because the derivation $Q_0 \vtr \dots \vtr
Q_i$ is of length $i$.  Now let $ Q_i \vtr Q_i \cup \{ \Phi \} = Q_{i+1}$ be the
derivation of the emptiness test.  If already $\Phi \in Q_i$ then $ Q_{i+1}
\subseteq \comp {\mcS_i}$ and we are done.

If $\Phi \not \in Q_i$, then

$Q_0\subseteq Q_i$ implies that $\Delta_G(\Phi,\Phi) \neq \emptyset$.

Since $\emptyset$ is an active state, we know that
$\emptyset \not \in Q_i$, and for $Q_i \vtr Q_{i+1}$ to be a
possible derivation of the emptiness test, $\Delta_G(\Phi,\Phi) = \exp {\Psi_1}
\times \dots \times \exp {\Psi_n} \neq\{ (\emptyset, \dots, \emptyset) \}$ must hold,
i.e., there must be a $\Psi_i \neq \emptyset$ such that $\exp {\Psi_i} \subseteq
Q_i \subseteq \comp {\mcS_i}$.  Hence $\pi_i \in \Phi$ and $\Psi_i = \{ \pi_i
\Diamond \} \cup \{ \pi \Box \mid \pi \in \Phi \mbox{ is a $\Box$-path}\}$.

Lemma~\ref{lem:inverse-calculus-for-propositional-closure} yields the existence
of a set of sequents $\mcT_i$ with $\mcS_i \vdstar \mcT$ containing
a sequent $\Lambda$ with $\Lambda \subseteq \Psi_i$. This sequent is either of
the form $\Lambda = \Gamma \Box, \pi_i \Diamond$ or $\Lambda = \Gamma \Box$ for some
$\Gamma \subseteq \Phi$. In the former case, \IC can use a $\Diamond$-inference
\begin{equation*}\label{eq:diamond-inference}
  \inference[$(\Diamond)$]{\Gamma \Box, \pi_i \Diamond}{\Gamma,\pi_i}
\end{equation*}
and in the latter case a $\Diamond^+$-inference 
\begin{equation*}\label{eq:diamond-plus-inference}
  \inference[$(\Diamond^+)$]{\Gamma \Box}{\Gamma, \pi_i}
\end{equation*}
to derive $S_0 \vdstar \mcS_i \vdstar \mcT \vdash \mcT \cup \{ \Gamma, \pi_i \} =
\mcS$ and $\Phi \subseteq \comp{ \Gamma, \pi_i}$ holds. \qed

\section{Optimizations}
\label{sec:opti}

Since the inverse calculus can be seen as an on-the-fly implementation of the
emptiness test, optimizations of the inverse calculus also yield optimizations
of the emptiness test. We use the connection between the two approaches to
provide an easier proof of the fact that the optimizations of \IC{} introduced
by Voronkov \cite{Voronkov-ToCL-2001} do not destroy completeness of the
calculus.

\subsection{Unreachable states / redundant sequents}

States that cannot occur on any run starting with an initial state have no effect 
on the language accepted by the automaton. We call such states
\emph{unreachable}. In the following, we will determine certain types of 
unreachable states.

\begin{definition}
  Let $\pi, \pi_1, \pi_2 \in\Pi_G$.
  \begin{itemize}
  \item The \emph{modal length} of $\pi$ is the number of occurrences of $\Box$
    and $\Diamond$ in $\pi$.
    
  \item $\pi_1, \pi_2 \in \Pi_G$ form a \emph{$\vee$-fork} if $\pi_1 = \pi
    {{\vee_l}} \pi_1'$ and $\pi_2 = \pi {{\vee_r}} \pi_2'$ for some $\pi, \pi_1',
    \pi_2'$.
    
  \item  $\pi_1,\pi_2$ are \emph{$\Diamond$-separated} if $\pi_1 =
    \pi_1' \Diamond \pi_1''$ and $\pi_2 = \pi_2' \Diamond \pi_2''$ such that
    $\pi_1', \pi_2'$ have the same modal length and $\pi_1' \neq \pi_2'$.
  \end{itemize}
\end{definition}

\begin{lemma}\label{lem:non-wellformed-sets}
  Let $\mfA_G$ be the formula automaton for a $\mathsf{K}$-formula $G$ in NNF
  and $\Phi \in Q$.   
  If $\Phi$ contains a $\vee$-fork, two $\Diamond$-separated paths, or two paths
  of different modal length, then $\Phi$ is unreachable.
\end{lemma}

The lemma shows that

we can remove such states from $\mfA_G$ without changing the accepted language.
Sequents containing a $\vee$-fork, two $\Diamond$-separated paths, or two paths
of different modal length represent only unreachable states, and are thus redunant,
i.e., inferences involving such sequents need not be considered.

\begin{definition}[Reduced automaton]
  Let $\bar Q$ be the set of states of $\mfA_G$ that contain a $\vee$-fork,
  two $\Diamond$-separated paths, or two paths of different modal length.  The
  \emph{reduced} automaton $\mfA'_G = (Q'_G, \Sigma_G, \exp { \{ \epsilon \} },
  \Delta'_G)$ is defined by

$$
    Q'_G := Q_G \setminus \bar Q\ \ \ \mbox{and}\ \ \
    \Delta'_G := \Delta_G \cap (Q'_G \times \Sigma_G \times Q'_G \times \dots \times Q'_G).
$$

\end{definition}

Since the states in $\bar Q$ are unreachable, 
$L(\mfA_G) = L(\mfA'_G)$. From now on, we consider 
$\mfA'_G$ and define $\comp \cdot$ relative to the states on $\mfA'_G$:
$\comp \Gamma = \{ \Phi \in Q'_G \mid \Gamma \subseteq \Phi \}$.

\subsection{$G$-orderings / redundant inferences}

In the following, the applicability of the propositional inferences of the
inverse calculus will be restricted to those where the affected paths are
maximal w.r.t.\  a total ordering of $\Pi_G$. In order to maintain completeness,
one cannot consider arbitrary orderings in this context.

  Two paths $\pi_1,\pi_3$ are \emph{brothers} iff there exists a \wedgevee-path
  $\pi$ such that $\pi_1 = \pi \wedgevee_l$ and $\pi_3 = \pi \wedgevee_r$ or
  $\pi_1 = \pi \wedgevee_r$ and $\pi_3 = \pi \wedgevee_l$.

\begin{definition}[$G$-ordering]
  Let $G$ be a $\mathsf{K}$-formula in NNF. A total ordering $\succ$ of $\Pi_G$
  is called a \emph{$G$-ordering} iff
  \begin{enumerate}
  \item $\pi_1 \succ \pi_2$ whenever
    \begin{enumerate}
    \item the modal length of $\pi_1$ is strictly greater than the modal length
      of $\pi_2$; or
    \item $\pi_1,\pi_2$ have the same modal length, the last symbol of $\pi_1$
      is $\wedgevee_*$, and the last symbol of $\pi_2$ is \boxdiamond; or
    \item  $\pi_1,\pi_2$ have the same modal length and $\pi_2$ is a prefix of $\pi_1$
    \end{enumerate}
  \item 
    There is no path between brothers, i.e., there exist no $G$-paths
    $\pi_1,\pi_2, \pi_3$ such that $\pi_1 \succ \pi_2 \succ \pi_3$ and
    $\pi_1,\pi_3$ are brothers.
  \end{enumerate}
\end{definition}  

For the example formula $G$ of Figure~\ref{fig:g-paths}, a $G$-ordering
$\succ$ can be defined by setting $\nu_9 \succ \nu_8 \succ \dots \succ \nu_1
\succ \nu_0$.
Voronkov \cite{Voronkov-ToCL-2001} shows that $G$-orderings exist for every 
$\mathsf{K}$-formula $G$ in NNF.

Using an arbitrary, but fixed $G$-ordering $\succ$, the applicability of the
propositional inferences is restricted as follows.

\begin{definition}[Optimized Inverse Calculus]
  For a sequent $\Gamma$ and a path $\pi$ we
  write $\pi \succ \Gamma$ iff $\pi \succ \pi'$ for every $\pi' \in \Gamma$.
  \begin{itemize}
  \item An inference $\inference[$({\wedge_*})$]{\Gamma, \pi {\wedge_*}}{\Gamma,
      \pi}$ \emph{respects} $\succ$ iff $\pi {\wedge_*} \succ \Gamma$.
  \item An inference $\inference[$(\vee)$]{\Gamma_l, \pi {\vee_l} & \Gamma_r,
      \pi {\vee_r}}{\Gamma_l, \Gamma_r, \pi}$ \emph{respects} $\succ$ iff $\pi
    {\vee_l} \succ \Gamma_l$ and $\pi {\vee_r} \succ \Gamma_r$.
  \item The $\Diamond$- and $\Diamond^+$-inferences always respect $\succ$.
  \end{itemize}
  The optimized inverse calculus \OIC works as \IC, but for each
  derivation $\mcS_0 \vdash \dots \vdash \mcS_k$ 
  the following \emph{restrictions} must hold:
  \begin{itemize}

  \item  For every step $\mcS_i \vdash \mcS_{i+1}$, the employed inference respects
    $\succ$, and

  \item $\mcS_i$ must not contain $\vee$-forks, $\Diamond$-separated paths, or paths of
               different modal length.

  \end{itemize}
\end{definition}

To distinguish derivations of \IC and \OIC, we will use the symbol $\vds$ in
derivations of $\OIC$.

In \cite{Voronkov-ToCL-2001}, correctness of \OIC is shown.

\begin{fact}[\cite{Voronkov-ToCL-2001}]
  \label{fact:optimised-calculus-correct}
  Let $G$ be a $\K$-formula in NNF and $\succ$ a $G$-ordering. Then $G$ is 
  unsatisfiable iff $\{ \epsilon \} \in \mcS_0^{\vds}$.
\end{fact}

Using the correspondence between the inverse method and the emptiness test of
$\mfA'_G$, we will now give an alternative, and in our opinion simpler, proof of
this fact.  Since \OIC is merely a restriction of \IC, soundness (i.e., the
if-direction of the fact) is immediate.

Completeness requires more work. In particular, the proof of
Lemma~\ref{lem:inverse-calculus-for-propositional-closure} needs to be
reconsidered since the propositional inferences are now restricted:
we must show that the $\wedgevee$-inferences employed in that proof respect
(or can be made to respect) $\succ$.

To this purpose, we will follow \cite{Voronkov-ToCL-2001} and introduce the
notion of $\succ$-compactness. For $\succ$-compact sets, we can be sure that all
applicable $\wedgevee$-inferences respect $\succ$.
To ensure that all the sets $\Upsilon_{i}$ constructed in the proof of 
Lemma~\ref{lem:inverse-calculus-for-propositional-closure} are $\succ$-compact,
we again follow Voronkov and employ a special selection strategy.

\begin{definition}[$\succ$-compact, $\select_\succ$]\label{def:compact}
  Let $G$ be a $\mathsf{K}$-formula in NNF and $\succ$ a $G$-ordering. An
  arbitrary set $\Phi \subseteq \Pi_G$ is \emph{$\succ$-compact} iff, for every
  \wedgevee-path $\pi \in \Phi$ that is not p.e.\ in $\Phi$, $\pi \wedgevee_*
  \succ \Phi$.
  
  The selection function $\select_\succ$ is defined as follows: if $\Phi$ is not
  p.e., then let $\{ \pi_1, \dots, \pi_m \}$ be the set of \wedgevee-paths that
  are not p.e.\ in $\Phi$. From this set, $\select_\succ$
  selects the path $\pi_i$ such that the paths $\pi_i \wedgevee_*$ are the two
  smallest elements in $\{ \pi_j \wedgevee_* \mid 1 \leq j \leq m \}$.
\end{definition}

The function $\select_\succ$ is well-defined because of Condition (2)
of $G$-orderings.  The definition of compact ensures that
$\wedgevee$-inferences applicable to not propositionally expanded sequents
respect $\succ$.

\begin{lemma}\label{lem:selection-enforces-compactness}
  Let $G$ be a $\mathsf{K}$-formula in NNF, $\succ$ a $G$-ordering, and
  $\select_\succ$ the selection function as defined above.

  Let $\Phi = \{ \epsilon \}$ or $\Phi = \Gamma \Box, \pi_i \Diamond$ with
  $\Box$-paths $\Gamma$ and a $\Diamond$-path $\pi$, all of equal modal length.
  If $\Tree_\Phi$, as defined in the proof of
  Lemma~\ref{lem:inverse-calculus-for-propositional-closure}, is generated using
  $\select_\succ$ as selection function, then every node $\Psi$ of $\Tree_\Phi$
  is $\succ$-compact.
\end{lemma}

\begin{proof}
  The proof is similar to the proof of Lemma~5.8.3 in \cite{Voronkov-ToCL-2001}.
  It is given by induction on the depths of the node $\Psi$ in the tree
  $\Tree_\Phi$. For the root $\Phi$ there are two possibilities.  If $\Phi = \{
  \epsilon \}$ and $\epsilon$ is a $\wedgevee$-path, then $\wedgevee_l$ and
  $\wedgevee_r$ have the same modal length as $\epsilon$ and $\wedgevee_* \succ
  \epsilon$ by Condition (1c) of $G$-orderings.  If $\Phi = \Gamma \Box, \pi_i
  \Diamond$ and $\pi \in \Phi$ is a $\wedgevee$-path, then $\pi \wedgevee_*
  \succ \Phi$ holds by Condition (1b) of $G$-orderings because the last symbol
  of every path in $\Phi$ is \boxdiamond.
  
  For the induction step, let $\Psi$ be a node in $\Tree_\Phi$ which we have
  already shown to be $\succ$-compact. We show that then also its successor
  nodes (if any) are $\succ$-compact.
  \begin{itemize}
  \item If $\Psi$ is an $\wedge$-node with selected $\wedge$-path $\pi \in
    \Psi$, then the successor node of $\Psi$ is $\Psi' = \Psi, \pi {\wedge_l}, \pi
    {\wedge_r}$. Let $\pi' \in \Phi'$ be a $\wedgevee$-path that is not p.e.\ in
    $\Phi'$. There are two possibilities:
    \begin{itemize}
    \item $\pi' = \pi {\wedge_*}$. In this case, since $\pi {\wedge_*} \wedgevee_*
      \succ \pi {\wedge_*}$ by Condition (1c) of $G$-orderings and $\pi {\wedge_*}
      \succ \Psi$, $\pi' \wedgevee_* \succ \Psi'$ holds.
    \item $\pi' \neq \pi {\wedge_*}$.  Then, $\pi' \in \Psi$ and $\pi' \neq \pi$
      holds because $\pi$ is p.e.\ in $\Psi'$. Since $\Psi$ is $\succ$-compact,
      $\psi' \wedgevee_* \succ \nu$ for every $\nu \in \Psi$.
      It remains to show that $\pi' \wedgevee_* \succ \pi \wedgevee_*$, which
      follows from the fact that $\pi$ was selected by $\select_\succ$.
    \end{itemize}
  \item If $\Psi$ is an $\vee$-path and the selected $\vee$-path is $\pi \in
    \Psi$, then, w.o.l.g., $\Phi = \Psi, \pi {\vee_l}$. The same arguments as
    before apply. \qed
  \end{itemize}
\end{proof}

Given this lemma, it is easy to show that the construction employed in the proof of
Lemma~\ref{lem:inverse-calculus-for-propositional-closure} also works for \OIC,
provided that we restrict the set $\Phi$ as in
Lemma~\ref{lem:selection-enforces-compactness}:

\begin{lemma}\label{lem:opt-inverse-calculus-for-propositional-closure}
  Let $\Phi = \{ \epsilon \}$ or $\Phi = \Gamma \Box, \pi_i \Diamond$ with
  $\Box$-paths $\Gamma$ and a $\Diamond$-path $\pi$ all of equal modal length
  and $\mcS$ a set of sequents such that $\exp \Phi \subseteq \comp \mcS$. Then
  there exists a set of sequents $\mcT$ with $\mcS \vdsstar \mcT$ such that there
  exists $\Lambda \in \mcT$ with  $\Lambda \subseteq \Phi$.
\end{lemma}


\begin{proof}We use the same construction as in the proof of
  Lemma~\ref{lem:inverse-calculus-for-propositional-closure}, but the special
  selection function $\select_\succ$ as above.  From
  Lemma~\ref{lem:selection-enforces-compactness} we have that all nodes
  $\Upsilon_i$ in $\Tree_\Phi$ are $\succ$-compact.  All we have to do is to
  make sure that the employed inferences respect $\succ$. We refer to the
  inferences by number assigned to them in the proof of
  Lemma~\ref{lem:inverse-calculus-for-propositional-closure}.
  \begin{itemize}
  \item [(\ref{eq:and-inference})] Since $\Upsilon_{i+1}$ is compact and $\pi
    \in \Upsilon_{i+1}$ is not p.e.\ in $\Upsilon_{i+1}$, $\pi {\wedge_l} \succ
    \Upsilon_{i+1}$ and hence $\pi {\wedge_l} \succ \Gamma$ because $\Gamma
    \subseteq \Upsilon_{i+1}$.
  \item [(\ref{eq:and-inference-2})] W.l.o.g., assume $\pi {\wedge_l} \succ \pi
    {\wedge_r}$. (If this is not the case, then reverse the order of the two
    inferences.) Since $\Upsilon_{i+1}$ is compact, $\Gamma \subseteq
    \Upsilon_{i+1}$ and $\pi \in \Upsilon_{i+1}$ is not p.e., $\pi {\wedge_l}
    \succ \Gamma$ holds as well as $\pi {\wedge_l} \succ \pi {\wedge_r}$. Also $\pi
    {\wedge_r} \succ \Gamma$ holds, which means that both inferences respect
    $\succ$.
  \item [(\ref{eq:or-inference})] Since $\Upsilon_{i+1}$ is compact and $\pi \in
    \Upsilon_{i+1}$ is not p.e.\ we have $\pi {\vee_*} \succ \Upsilon_{i+1}$ and
    since both $\Gamma_l$ and $\Gamma_r$ are subsets of $\Upsilon_{i+1}$, also
    $\pi {\vee_l} \succ \Gamma_l$ and $\pi {\vee_r} \succ \Gamma_r$ holds. \qed
  \end{itemize}
\end{proof}

\paragraph{Alternative Proof of Fact~\ref{fact:optimised-calculus-correct}.}
  
As mentioned before, soundness (the if-direction) is immediate.  For the
only-if-direction, if $G$ is not satisfiable, then $L(\mfA'_G) = \emptyset$ and 
there is a set of states $Q$ with $Q_0 \vtrstar Q$ and $\exp {\{ \epsilon \}} \subseteq Q$. 
Using Lemma~\ref{lem:opt-inverse-calculus-for-propositional-closure} we show that
there is a derivation of \OIC that simulates this derivation, i.e., there is a
set of sequents $\mcS$ with $\mcS_0 \vdsstar \mcS$ and $Q \subseteq \comp
\mcS$.
  
The proof is by induction on the length $m$ of the derivation $Q_0 \vtr \dots \vtr Q_m
= Q$ and is totally analogous to the proof of
Theorem~\ref{theo:emptiness-test-to-inv-calc}. The base case is
Lemma~\ref{lem:dead-states-equal-axioms}, which also holds for \OIC and the
reduced automaton. The induction step uses 
Lemma~\ref{lem:opt-inverse-calculus-for-propositional-closure} instead of
Lemma~\ref{lem:inverse-calculus-for-propositional-closure}, but this is the only
difference.
  
Hence, $Q_0 \vtrstar Q$ and $\exp {\{ \epsilon \}} \subseteq Q$ implies that
there exist a derivation $\mcS_0 \vdsstar \mcS$ such that 
$\exp {\{ \epsilon \}} \subseteq \comp {\mcS}$.
Lemma~\ref{lem:opt-inverse-calculus-for-propositional-closure} yields a
derivation $\mcS \vdsstar \mcT$ with $\{ \epsilon \} \in \mcT \subseteq
\mcS_0^{\vds}$. \qed

\section{Global axioms}

\def\axioms{\textit{ax}}

When considering satisfiability of $G$ w.r.t.\ the global axiom $H$,
we must take subformulae of $G$ and $H$ into account. We 
address subformulae using paths in $G$ and $H$.

\begin{definition}[$(G,H)$-Paths]
  For $\K$-formulae $G,H$ in NNF, the set of $(G,H)$-paths $\Pi_{G,H}$ is a subset
  of  $\{ \epsilon_G, \epsilon_H\}{\cdot}\{ {\vee_l},
  {\vee_r}, {\wedge_l}, \wedge_r, \Box, \Diamond \}^*$. The set $\Pi_{G,H}$ and
  the subformula $(G,H)|_\pi$ of $G,H$ addressed by a path $\pi \in \Pi_{G,H}$ are
  defined inductively as follows:
  \begin{itemize}
  \item $\epsilon_G \in \Pi_{G,H}$ and $(G,H)|_{\epsilon_G} = G$,\ \ and\ \
        $\epsilon_H \in \Pi_{G,H}$ and $(G,H)|_{\epsilon_H} = H$
  \item if $\pi \in \Pi_{G,H}$ and $(G,H)|_\pi = F_1 \wedge F_2$ then $\pi
    {\wedge_l}, \pi {\wedge_r} \in \Pi_{G,H}$, $(G,H)|_{\pi {\wedge_l}} = F_1$,
    $(G,H)|_{\pi {\wedge_r}} = F_2$, and $\pi$ is called \emph{$\wedge$-path}.
  \item The other cases are defined analogously (see also Definition~\ref{def:g-paths}). 
  \item $\Pi_{G,H}$ is the smallest set that satisfies the previous
    conditions.
  \end{itemize}  
\end{definition}

The definitions of \emph{p.e.} and \emph{clash} are extended to subsets of
$\Pi_{G,H}$ in the obvious way, with the \emph{additional requirement} that, for
$\Phi \neq \emptyset$ to be p.e., $\epsilon_H \in \Phi$ must hold. This
additional requirement enforces the global axiom.

\begin{definition}[Formula Automaton with Global Axioms]\label{def:axiom-automaton}
  For $\mathsf{K}$-for\-mu\-lae $G,H$ in NNF, let $\{ \pi_1, \dots, \pi_n \}$ be an
  enumeration of the $\Diamond$-paths in $\Pi_{G,H}$.

  The $n$-ary looping automaton $\mfA_{G,H}$ is defined by 
  \[
  \mfA_G := (Q_{G,H},  \Sigma_{G,H}, \exp{\{ \epsilon_G \}}, \Delta_{G,H}),
  \]
  where $Q_{G,H} := \Sigma_{G,H} := \{ \Phi \in \Pi_{G,H} \mid \Phi \text{ is
    p.e.} \}$ and the transition relation $\Delta_{G,H}$ is defined as for the
  automaton $\mfA_G$ in Definition~\ref{def:automaton}.
\end{definition}

\begin{theorem}\label{lem:axioms-emptiness-and-non-satisfiability}
  $G$ is satisfiable w.r.t.\ the global axiom
  $H$ iff $L(\mfA_{G,H}) \neq \emptyset$.
\end{theorem}

\begin{proof}
  The proof is totally analogous to the proof of
  Theorem~\ref{lem:emptiness-and-non-satisfiability}. We use the same
  constructions for both directions.
  
  Let $\{\pi_1, \dots, \pi_n\}$ be an enumeration of the $\Diamond$-paths in
  $\Pi_{G,H}$.
  
  For the \emph{if}-direction let $L(\mfA_{G,H}) \neq \emptyset$, $t,r : [n]^*
  \rightarrow \{ \Phi \subseteq \Pi_{G,H} \mid \Phi \text{ is p.e.} \}$ a tree
  that is accepted by $\mfA_{G,H}$ and a corresponding run of $\mfA_{G,H}$. By
  construction of $\mfA_{G,H}$, $t(w) = r(w)$ for every $w \in [n]^*$.  We
  construct a Kripke model $\mcM = (W,R,V)$ from $t$ by setting
  \begin{align*}
    W & = \{ w \in [n]^* \mid t(w) \neq \emptyset \}\\
    R & = \{ (w,wi) \in W \times W \mid i \in [n] \}\\
    V & = \lambda P . \{ p \in W \mid \exists \pi \in t(w) . (G,H)|_\pi = P \} \quad
    \text{ for all propositional atoms $P$ }
  \end{align*}

  \begin{claim}
    For all $w \in W$, if $\pi \in t(w)$ then $\mcM,w \models (G,H)|_\pi$.
  \end{claim}
  
  \noindent \textit{Proof of the claim.} The claim is proved by induction on the
  structure of \K-formulae.  Let $w \in W$ be a world and $\pi \in \Pi_G$ be a
  path such that $\pi \in t(w)$. 
  \begin{itemize}
  \item if $(G,H)|_\pi = P$ is a propositional atom and $w \in W$, then $w \in
    V(P)$ and hence $\mcM,w \models (G,H)|_\pi$.
  \item if $(G,H)|_\pi = \neg P$ is a negated propositional atom, then, since
    $t(w)$ is clash free, there is no $\pi' \in \Pi_{G,H}$ such that
    $(G,H)|_{\pi'} = P$. Thus, $w \not \in V(P)$ and hence $\mcM,w \models \neg
    P$.
  \item if $(G,H)|_\pi = F_1 \wedge F_2$ then $\pi$ is an $\wedge$-paths, and
    since $t(w)$ is p.e., $\{ \pi {\wedge_l}, \pi {\wedge_r}\} \subseteq t(w)$.
    By induction, $\mcM, w \models (G,H)|_{\pi{\wedge_*}}$ and hence $\mcM,w
    \models (G,H)|_\pi$.
  \item if $(G,H)|_\pi = F_1 \vee F_2$ then $\pi$ is an $\vee$-paths, and since
    $t(w)$ is p.e., $\{ \pi {\vee_l}, \pi {\vee_r}\} \cap t(w) \neq \emptyset$.
    By induction, $\mcM, w \models (G,H)|_{\pi{\vee_l}}$ or $\mcM, w \models
    (G,H)|_{\pi{\vee_r}}$ and hence $\mcM,w \models (G,H)|_\pi$.
  \item if $(G,H)|_\pi = \Diamond F$ then $\pi$ is a $\Diamond$-path and,
    w.o.l.g., assume $\pi = \pi_i$. Since $\pi_i \in r(w)$, $\pi_i \Diamond \in
    r(wi) = t(wi)$ holds and hence $wi \in W$ and $(w,wi) \in R$. By induction,
    we have that $\mcM, wi \models (G,H)|_{\pi_i \Diamond}$ and hence $\mcM, w
    \models (G,H)|_{\pi_i}$.
  \item if $(G,H)|_\pi = \Box F$ and $(w,w') \in R$ then $w' = wi$ for some $i
    \in [n]$ and $t(wi) \neq \emptyset$ holds and by construction of
    $\mfA_{G,H}$, this implies $\pi \Box \in r(wi) = t(wi)$. By induction, this
    implies $\mcM, wi \models (G,H)|_{\pi \Box}$ and since $wi = w'$ and $w'$
    has been chosen arbitrarily, $\mcM, w \models (G,H)|_\pi$.
  \end{itemize}
  
  This finishes the proof of the claim. Since $t(\epsilon) = r(\epsilon) \in
  \exp{ \{ \epsilon_G \} }$ and hence $\epsilon_G \in t(\epsilon)$, $\mcM,
  \epsilon \models (G,H)|_{\epsilon_G}$ and $G = (G,H)|_{\epsilon_G}$ is
  satisfiable.
  
  Also, since $t(w)$ is p.e., $\epsilon_H \in t(w)$ for
  every $w \in W$ and, by the claim, $\mcM, w \models H = (G,H)|_{\epsilon_H}$
  holds for every $w \in W$. Hence $G$ is satisfiable w.r.t.\ the global axiom
  $H$.
  
  For the \emph{only if}-direction, we first show an auxiliary claim: for a set
  $\Psi \subseteq \Pi_{G,H}$ we define $\mcM, w \models \Psi$ iff $\mcM, w
  \models (G,H)|_\pi$ for every $\pi \in \Psi$.

  \begin{claim}
    If $\Psi \subseteq \Pi_{G,H}$ and $w \in W$ such that $\mcM, w \models
    \Psi$, then there is a $\Phi \in \exp \Psi$ such that $\mcM, w \models
    \Phi$.
  \end{claim}
  
  \noindent \textit{Proof of the claim.} Let $\Psi \subseteq \Pi_{G,H}$ and $w
  \in W$ such that $\mcM, w \models \Psi$. 
  We will show how to construct an expansion of $\Psi$ with the desired
  property.  If $\Psi$ is already p.e., then $\Psi \in \exp \Psi$ and we are
  done.
  
  \begin{itemize}
  \item If $\Psi$ is not p.e.\ because $\epsilon_H \not \in \Psi$ then, because
    $\mcM, w \models H$, $\Psi' = \Psi \cup \{ \epsilon_H \}$ is a set with
    $\mcM, w \models \Psi$ that is ``one step closer'' to being p.e.\ than
    $\Psi$.  
  \item If $\Psi$ is not p.e.\ and $\epsilon_H \in \Psi$ then let $\pi \in
    \Psi$ be a \wedgevee-path that is not p.e.\ in $\Psi$.
    \begin{itemize}
    \item If $\pi$ is a $\wedge$-path then $(G,H)|_\pi = F_1 \wedge F_2$ and
      since $\mcM, w \models (G,H)|_\pi$, also $\mcM, w \models F_1 =
      (G,H)|_{\pi \wedge_l}$ and $\mcM, w \models F_2 = (G,H)|_{\pi \wedge_r}$.
      Hence $\mcM, w \models \Psi \cup \{ \pi{\wedge_l}, \pi{\wedge_r} \}$ and
      $\Psi' = \Psi \cup \{ \pi{\wedge_l}, \pi{\wedge_r}\}$ is a set with $\mcM,
      w \models \Psi'$ that is ``one step closer'' to being p.e.\ than $\Psi$.
    \item If $\pi$ is a $\vee$-path then $(G,H)|_\pi = F_1 \vee F_2$ and since
      $\mcM, w \models (G,H)|_\pi$, also $\mcM, w \models F_1 = (G,H)|_{\pi
        \vee_l}$ or $\mcM, w \models F_2 = (G,H)|_{\pi \vee_r}$. Hence $\mcM, w
      \models \Psi \cup \{ \pi{\vee_l}\}$ or $\mcM, w \models \Psi \cup \{
      \pi{\vee_r}\}$ and hence can obtain a set $\Psi'$ with $\mcM, w \models
      \Psi'$ that is again ``one step close'' to being p.e. than $\Psi$.
    \end{itemize}  
  \end{itemize}

  Restarting this process with $\Psi = \Psi'$ eventually yields an expansion
  $\Phi$ of the initial set $\Psi$ with $\mcM,w \models \Phi$, which proves the
  claim.
  
  Let $\mcM = (W,R,V)$ be a model for $G$ with $w \in W$ such that $\mcM, w
  \models G$. From $\mcM$ we construct a tree that is accepted by $\mfA_{G,H}$.
  Using this claim, we inductively define a tree $t$ accepted by $\mfA_{G,H}$.
  To this purpose, we also inductively define a function $f : [n]^* \rightarrow
  W$ such that, if $\mcM, f(p) \models t(p)$ for all $p$.
  
  We start by setting $f(\epsilon) = w$ for a $w \in W$ with $\mcM, w \models
  G$.  and $t(\epsilon) = \Phi$ for a $\Phi \in \exp { \{ \epsilon \} }$ such
  that $\mcM, w \models \Phi$. From the claim we have that such a set $\Phi$
  exists because $\mcM, w \models G = (G,H)|_\epsilon$.
  
  If $f(p)$ and $t(p)$ are already defined, then, for $i \in [n]$, we define
  $f(pi)$ and $t(pi)$ as follows:
  \begin{itemize}
  \item if $\pi_i \in t(p)$ then $\mcM, f(p) \models (G,H)|_{\pi_i}$ and hence
    there is a $w' \in W$ such that $(f(p),w') \in R$ and $\mcM, w' \models
    (G,H)|_{\pi_i \Diamond}$. If $\pi \in t(p)$ is a $\Box$-path, then also
    $\mcM, w' \models (G,H)|_{\pi\Box}$ holds. Hence $\mcM, w' \models \{ \pi_i
    \Diamond \} \cup \{ \pi \Box \mid \pi \in t(p) \text{ is a $\Box$-path }
    \}$.  We set $f(pi) = w'$ and $t(pi) = \Phi$ for a $\Phi \in \exp {\{ \pi_i
      \Diamond \} \cup \{ \pi \Box \mid \pi \in t(p) \text{ is a $\Box$-path }
      \}}$ with $\mcM, w' \models \Phi$, which exist by the claim.
  \item if $\pi_i \not \in t(p)$, then we set $f(pi) = w$ for an arbitrary $w
    \in W$ and $t(pi) = \emptyset$
  \end{itemize}
  In both cases, we have define $f(pi)$ and $t(pi)$ such that $\mcM, f(pi)
  \models t(pi)$. It is easy to see that $t$ is accepted by $\mfA_{G,H}$ with
  the run $r = t$. Hence $L(\mfA_{G,H}) \neq \emptyset$ which is what we needed
  to show. \qed
\end{proof}

\begin{definition}[The Inverse Calculus w.\ Global Axiom]
  Let $G,H$ be \K- formula in NNF and $\Pi_{G,H}$ the set of paths of $G,H$.
  Sequents are subsets of $\Pi_{G,H}$, and operations on sequents are defined as before.
  
  In addition to the inferences from Figure \ref{fig:inv-calc}, the inverse
  calculus for $G$ w.r.t.\ the global axiom $H$, \AIC, employs the inference
  \[
  \inference[$(\axioms)$]{\Gamma, \epsilon_H}{\Gamma}.
  \]
\end{definition}

From now on, $\comp \cdot$ is defined w.r.t.\ the states of $\mfA_{G,H}$, i.e.,
$\comp \Gamma := \{ \Phi \in Q_{G,H} \mid \Gamma \subseteq \Phi \}$.

\begin{theorem}[\AIC and the emptiness test for $\mfA_{G,H}$ simulate each other]%
\label{theo:axioms-inv-calc-to-emptiness-test}%
\label{theo:axioms-emptiness-test-to-inv-calc}%
Let $\vda$ denote derivation steps of \AIC, and $\vtr$ derivation steps of the emptiness
test for $\mfA_{G,H}$. 
  \begin{enumerate}
  \item Let $Q\subseteq Q_{G,H}$ be a set of states such that $Q_0 \vtrstar Q$.
    Then there exists a set of sequents $\mcS$ with
    $\mcS_0 \vdastar \mcS \text{ and } Q \subseteq \comp \mcS$.
  \item  Let $\mcS$ be a set of sequents
    such that $\mcS_0 \vdastar \mcS$. Then there exists a set of states $Q
    \subseteq Q_G$ with
    $Q_0 \vtrstar Q \text{ and } \comp \mcS \subseteq Q$.
  \end{enumerate}
\end{theorem}

Lemma~\ref{lem:dead-states-equal-axioms},
\ref{lem:propositional-inferences-for-free}, and
\ref{lem:modal-inference-simulated-by-emptiness-test}, restated for $\mfA_{G,H}$
and \AIC, can be shown as before.

The following lemma deals with the $\axioms$-inference of \AIC.

\begin{lemma}\label{lem:axioms-inference-for-free}
  Let $\mcS \vtr \mcT$ be a derivation of \AIC that employs an
  \axioms-inference. Then $\comp \mcS = \comp \mcT$.
\end{lemma}

\begin{proof}
  Let $\mcT = \mcS \cup \{ \Gamma \}$ with $\{ \Gamma, \epsilon_H \} \in \mcS$.
  Then we know that $\inference[$(\axioms)$]{\Gamma, \epsilon_H}{\Gamma}$.
  $\comp \mcT = \comp \mcS \cup \comp \Gamma$. Since $\mcS \subseteq \mcT$,
  $\comp \mcS \subseteq \comp \mcT$ holds immediately. If $\Phi \in \comp
  \Gamma$, then, since $\Phi$ is p.e., $\epsilon_H \in \Phi$ and $\Phi \in \comp
  { \Gamma, \epsilon_H } \subseteq \comp \mcS$. \qed
\end{proof}

The proof of Theorem~\ref{theo:axioms-inv-calc-to-emptiness-test}.2 is now analogous to
the proof of Theorem~\ref{theo:inv-calc-to-emptiness-test}.2.

For the proof of Theorem~\ref{theo:axioms-inv-calc-to-emptiness-test}.1,
Lemma~\ref{lem:inverse-calculus-for-propositional-closure} needs to be re-proved
because the change in the definition of p.e.\ now also implies that $\epsilon_H
\in \Phi$ holds for every set $\Phi \in \exp \Psi$ for any $\Psi \neq \emptyset$
(see Lemma~\ref{lem:axioms-inverse-calculus-for-propositional-closure}). This is
where the new inference \axioms{} comes into play.  In all other respects, the
proof of Theorem~\ref{theo:axioms-inv-calc-to-emptiness-test}.1 is analogous to
the proof of Theorem~\ref{theo:inv-calc-to-emptiness-test}.1.

\begin{lemma}\label{lem:axioms-inverse-calculus-for-propositional-closure}
  Let $\Phi \subseteq \Pi_G$ a set of paths and $\mcS$ a set of sequents such
  that $\exp \Phi \subseteq \comp \mcS$. Then there exists a set of sequents
  $\mcT$ with $\mcS \vdastar \mcT$ such that there exists $\Lambda \in \mcT$
  with $\Lambda \subseteq \Phi$.
\end{lemma}

\begin{proof}
  If $\epsilon_H \in \Phi$ than we can use the same construction used in the
  proof of Lemma~\ref{lem:inverse-calculus-for-propositional-closure} to 
  construct the set $\mcT$ such that $\mcS \vdastar \mcT$ and there is a
  $\Lambda \in \mcT$ with $\Lambda \subseteq \Phi$. 
  
  If $\epsilon_H \not \in \Phi$, then set $\Psi = \Phi,\epsilon_H$ and again use
  the construction from the proof of Lemma
  ~\ref{lem:inverse-calculus-for-propositional-closure} to construct a set
  $\mcT$ such that $\mcS \vdastar \mcT$ and there is a $\Lambda \in \mcT$ with
  $\Lambda \subseteq \Psi$. If $\epsilon_H \not \in \Lambda$ then we are done
  since then also $\Lambda \subseteq \Phi$. If $\Lambda = \Gamma, \epsilon_H$
  for some $\Gamma$ with $\epsilon_H \not \in \Gamma$, then $\Gamma \subseteq
  \Phi$ and $\mcT \vda \mcT \cup \{ \Gamma \}$ can be derived by \AIC using the
  inference $\inference[$(\axioms)$]{\Gamma, \epsilon_H}{\Gamma}$. \qed
\end{proof}

\begin{corollary}
\AIC yields an \EXPTIME{} decision procedure for satisfiability w.r.t.\ global
axioms in \K.
\end{corollary}

The following algorithm yields the desired procedure:

\begin{algorithm}\label{alg:inverse-calculus}
  Let $G, H$ be \K-formulae in NNF. To test satisfiability of $G$ w.r.t.\ $H$,
  calculate $\mcS_0^{\vda}$. If $\{\emptyset, \{ \epsilon_G \}\} \cap \mcS_0^{\vda} \neq \emptyset$, 
  then answer ``not satisfiable,'' and ``satisfiable'' otherwise.
\end{algorithm}

Correctness of this algorithm follows from Theorem~\ref{lem:axioms-emptiness-and-non-satisfiability} 
and~\ref{theo:axioms-inv-calc-to-emptiness-test}. If $G$ is not satisfiable w.r.t.\ $H$,
then $L(\mfA_{G,H}) = \emptyset$, and there exists a set
of states $Q$ with $Q_0 \vtrstar Q$ and $\exp {\{ \epsilon_G \}} \subseteq Q$. 
Thus, there exists a set of sequents
$\mcS$ with $\mcS_0 \vdastar \mcS$ such that $Q \subseteq \comp {\mcS}$.
With (the appropriately reformulated) Lemma~\ref{lem:inverse-calculus-for-propositional-closure} 
there exists a
set of sequents $\mcT$ with $\mcS \vdastar \mcT$ such that there is a sequent
$\Lambda \in \mcT$ with $\Lambda \subseteq \{ \epsilon_G \}$. 
Consequently, $\Lambda = \emptyset$ or $\Lambda = \{ \epsilon_G \}$.

Since $\mcS_0 \vdastar\mcS_0^{\vda}$,  there exists a set of (inactive) states $Q$ such
that $Q_0 \vtrstar Q$ and $\comp{\mcS_0^{\vda}} \subseteq Q$. 
Since $\exp{\{\epsilon_G\}}\subseteq\comp{\{\epsilon_G\}}\subseteq\comp{\emptyset}$,
we know that
$\{\emptyset, \{ \epsilon_G \}\} \cap \mcS_0^{\vda} \neq \emptyset$ implies
$\exp{\{\epsilon_G\}}\subseteq Q$. Consequently, $L(\mfA_{G,H}) = \emptyset$
and thus $G$ is not satisfiable w.r.t.\ $H$.

For the complexity, note that there are only exponentially many sequents. Consequently, it
is easy to see that the saturation process that leads to $\mcS_0^{\vda}$ can be
realized in time exponential in the size of the input formulae.

\section{Future Work}

There are several interesting directions in which to continue this work.  First,
satisfiability in \K{} (without global axioms) is \textsc{PSpace}-complete
whereas the inverse method yields only an \EXPTIME-algorithm. Can suitable
optimizations turn this into a \textsc{PSpace}-procedure?  Second, can the
optimizations considered in Section~\ref{sec:opti} be extended to the inverse
calculus with global axioms?  Third, Voronkov considers additional
optimizations. Can they also be handled within our framework?  Finally, can the
correspondence between the automata approach and the inverse method be used to
obtain inverse calculi and correctness proofs for other modal or description
logics?

\bibliographystyle{plain}
\bibliography{strings,mybib,proceedings,aux}

\end{document}